\documentclass[11pt]{article}
\pdfoutput=1
\usepackage{jheppub}

\usepackage{color}
\usepackage{amsmath}
\usepackage{pifont}
\usepackage{bbold}

\usepackage{bbm}
\usepackage{verbatim}   
\usepackage{subfigure}
\usepackage{acronym}

\usepackage{amsfonts}
\usepackage{amssymb}
\usepackage{mathrsfs}
\usepackage{graphicx}
\usepackage{multirow}
\usepackage{slashed}

\graphicspath{{./fig/}}

\newcommand{\eg}{{\it e.g.}}          
                   
          \newcommand{\cm}{{\mathrm {cm}}}

 \newcommand{\MeV}{{\mathrm {MeV}}}
  \newcommand{\GeV}{{\mathrm {GeV}}}
   \newcommand{\TeV}{{\mathrm {TeV}}}

\def\inv{^{\raise.15ex\hbox{${\scriptscriptstyle -}$}\kern-.05em 1}}
\def\lbar{{\lower.35ex\hbox{$\mathchar'26$}\mkern-10mu\lambda}} 

\def\to{\rightarrow}

\let\<=\langle
\let\>=\rangle

\let\+=\uparrow

\newcommand{\newc}{\newcommand}
\newc{\gsim}{\lower.7ex\hbox{$\;\stackrel{\textstyle>}{\sim}\;$}}
\newc{\lsim}{\lower.7ex\hbox{$\;\stackrel{\textstyle<}{\sim}\;$}}

\newcommand{\gev}{\ensuremath{\text{ GeV}}}

\def\beq{\begin{equation}}
\def\eeq{\end{equation}}
\def\bea{\begin{eqnarray}}
\def\eea{\end{eqnarray}}

\newcommand{\ifb}{{{\text{ fb}}^{-1}}}

\newcommand{\MET}{\ensuremath{E_{T}^{\mathrm{miss}}}}

\begin{document}


%

\title{Auto-Concealment of Supersymmetry in Extra Dimensions}

\author[a]{Savas Dimopoulos,}
\emailAdd{savas@stanford.edu}

\author[a,b]{Kiel Howe,}
\emailAdd{howek@stanford.edu}

\author[c,a]{John March-Russell,}
\emailAdd{jmr@thphys.ox.ac.uk}

\author[c,d]{James Scoville}
\emailAdd{james.scoville@physics.ox.ac.uk}

\affiliation[a]{Stanford Institute for Theoretical Physics, Department of Physics,\\
Stanford University, Stanford, CA 94305, USA}
\affiliation[b]{SLAC National Accelerator Laboratory \\ Menlo Park, CA  94025 USA}
\affiliation[c]{Rudolf Peierls Centre for Theoretical Physics,
University of Oxford\\
1 Keble Road, Oxford,
OX1 3NP, UK}

\affiliation[d]{United States Air Force Institute of Technology\\Wright-Patterson Air Force Base, OH 45433, USA}

\abstract{In supersymmetric (SUSY) theories with extra dimensions the visible energy in sparticle decays can be significantly
reduced and its energy distribution broadened, thus significantly weakening the present collider limits on SUSY.  The mechanism
applies when the lightest supersymmetric particle (LSP) is a bulk state---\eg\  a bulk modulino, axino,
or gravitino---the size of the extra dimensions $\gsim 10^{-14}~\cm$, and for a broad variety of visible sparticle spectra.  In such
cases the lightest ordinary supersymmetric particle (LOSP), necessarily a brane-localised state, decays to the Kaluza-Klein (KK)
discretuum of the LSP.  This dynamically realises the compression mechanism for hiding SUSY as
decays into the more numerous heavier KK LSP states are favored.  We find LHC limits on right-handed slepton LOSPs evaporate, while LHC limits on stop LOSPs weaken to $\sim350\div410~\GeV$ compared to $\sim700~\GeV$ for a stop decaying to a massless LSP.  
Similarly, for the searches we consider, present limits on direct production of degenerate first and second generation squarks drop
to $\sim 450~\GeV$ compared to $\sim800~\GeV$ for a squark decaying to a massless LSP. 
Auto-concealment typically works for a fundamental gravitational scale of $M_*\sim 10 \div 100~\TeV$, a scale sufficiently high that traditional searches for signatures
of extra dimensions are mostly avoided.  If superpartners are discovered, their prompt, displaced, or stopped decays can also provide new
search opportunities for extra dimensions with the potential to reach $M_*\sim 10^9~\GeV$.
This mechanism applies more generally than just SUSY theories, pertaining
to any theory where there is a discrete quantum number shared by both brane and bulk sectors.
}

\maketitle


\section{Introduction}
\label{intro}

Discovery of supersymmetry (SUSY) at a hadron collider, such as the LHC, requires distinguishing SUSY signals from the large Standard Model (SM) backgrounds that are present.  Often this involves using the large missing transverse energy ($\MET$) or large visible energy present in a SUSY event.  Utilising such discriminators, LHC limits on
SUSY have significantly encroached into the region of parameter space consistent with natural electroweak symmetry breaking, with typical implied tuning of the order of 1\% or less \cite{Arvanitaki:2013yja,Gherghetta:2012gb,Hardy:2013ywa,Feng:2013pwa} weakening the case for low-scale SUSY as a solution to the hierarchy problem.\footnote{In Refs.\cite{MNSUSY,Garcia:2014lfa,moreMNSUSY} a class of SUSY models which are fully natural at present LHC limits were studied.  These involve extra dimensions and in fact can incorporate the mechanism discussed in this paper, though this fact was neither emphasised nor used in these studies.}  

In this paper we present a mechanism by which SUSY signals at a hadron collider are dynamically degraded.  We consider a framework in which the SM particles
and their SUSY-partners live on a brane that is embedded in a (flat) $4+d$-dimensional supersymmetric bulk whose dimensions are bigger than $\sim 10^{-14}\cm \sim 1/({\rm few}~\GeV)$. SUSY breaking is felt softly on the brane, and the MSSM superpartners may be produced at colliders\footnote{In contrast, studies of brane-worlds with supersymmetric bulks have focused mostly on the case that SUSY is realized only non-linearly on the brane \cite{Atwood:2000au,Burgess:2004yq,Matias:2005gg,Cicoli:2011yy,Antoniadis:2001pt,Dudas:2000nv,Klein:2002vu,Antoniadis:2004uk}.}. As we will show, many realizations of this scenario have additional light bulk states that are associated with SUSY breaking or additional sequestered sectors. In such cases, the lightest R-parity odd sparticle, the `bulk LSP', will propagate in the $4+d$ extra dimensions, and the lightest ordinary-sector SUSY particle (LOSP) will decay to this state. 

\begin{figure}[t]
  \centering
  \includegraphics[width=.7\columnwidth]{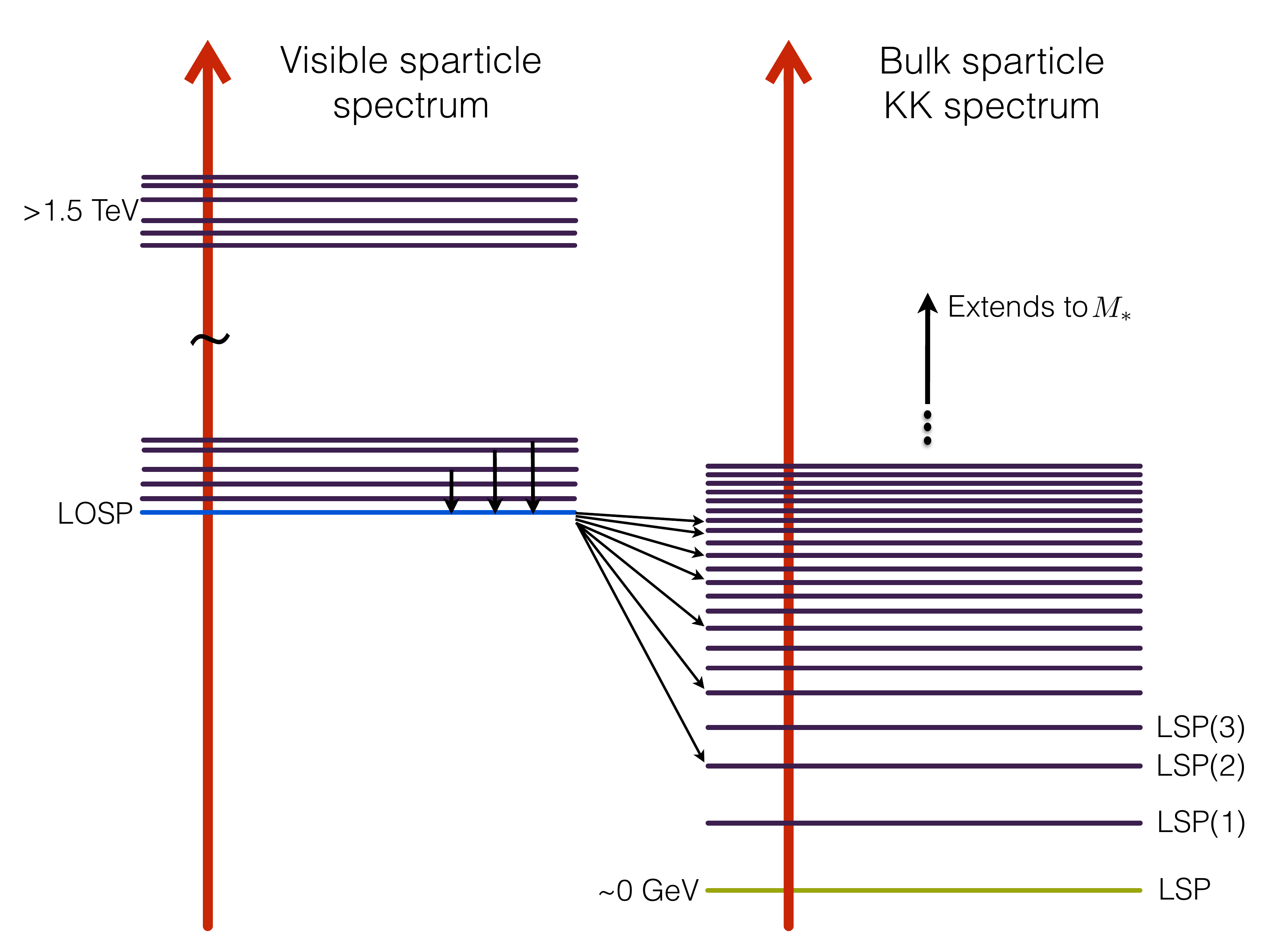}
  \caption{Schematic representation of the basic idea behind the auto-concealment mechanism, in which  the LSP is a bulk state propagating in $d\geq 1$ extra dimensions.  The visible sparticle spectrum has a lightest state, the LOSP, which decays promptly
  to the full tower of KK excitations of the LSP.   As the spectral density of KK excitations behaves as $\sim m_n^{d-1}$ (as a function of
  the KK mass, $m_n$), decays to the heavier KK states are favored, dynamically realising the compressed spectrum mechanism of hiding SUSY with reduced $\MET$ and visible
  energy. As the masses of the KK tower of the LSP extend from $\sim 0~\GeV$ to the underlying gravitational scale $M_*$ the LOSP mass is automatically within this tower without additional tuning.
Transitions from visible sector to the bulk sector are prompt if $M_*$ is not too high depending on the nature of the bulk LSP.   In the case that transitions are not prompt, the auto-concealment mechanism no longer functions, but instead the decays of the LOSP can provide a powerful search method for extra dimensions.}
    \label{fig:basicidea}
\end{figure}

Couplings between bulk and brane states are necessarily higher dimensional operators, and if the  fundamental  scale, $M_*$, is not too high, decays of the LOSP can occur on collider timescales. From a 4d perspective, the LOSP decays to a distribution of KK modes of the bulk LSP of mass $m_n$ with bulk phase space factor $\sim m_n^{d-1}$. This favors decays to the heaviest KK states, thus suppressing both visible energy and $\MET$ in the decay, and so, as we will argue in detail, severely weakening LHC limits on SUSY for certain classes of visible sparticle spectra. The basic mechanism is illustrated in Figure~\ref{fig:basicidea}.

Specifically, we show that two-body decays of the brane-localized LOSP of mass $M$ to a SM state and a bulk LSP are typically dominated by decays to bulk KK modes with masses $m_n\gtrsim0.4M\div 0.8 M$ depending on the nature of the coupling and the dimension of the bulk.  This leads automatically to signatures similar to a compressed spectrum, where super-partners with large production cross sections are concealed if they decay to a nearly degenerate invisible LSP\footnote{Ref.~\cite{Alves:2013wra} provides another example of a theory that dynamically reduces missing and visible energy, reproducing signatures similar to compressed spectra.} \cite{Izaguirre:2010nj,LeCompte:2011fh,LeCompte:2011cn,Dreiner:2012gx,Bhattacherjee:2012mz,Drees:2012dd,Belanger:2012mk,Bhattacherjee:2013wna}. Cascade decays that produce a highly boosted LOSP are not as effectively hidden, but nonetheless we find that a variety of motivated and potentially low-fine-tuned spectra are successfully auto-concealed.  In this work, we focus primarily on limits from searches for prompt decays, which restricts $M_*$ from above depending on the nature of the bulk LSP (modulino, axino, gaugino, gravitino) and the identity of the LOSP.   For higher scales of $M_*$, searches for displaced vertices and out-of-time stopped decays become relevant, and their sensitivity is also likely to be affected, though a study of this possibility is beyond the scope of this work.

\begin{figure}[t]
  \centering
  \includegraphics[width=.7\columnwidth]{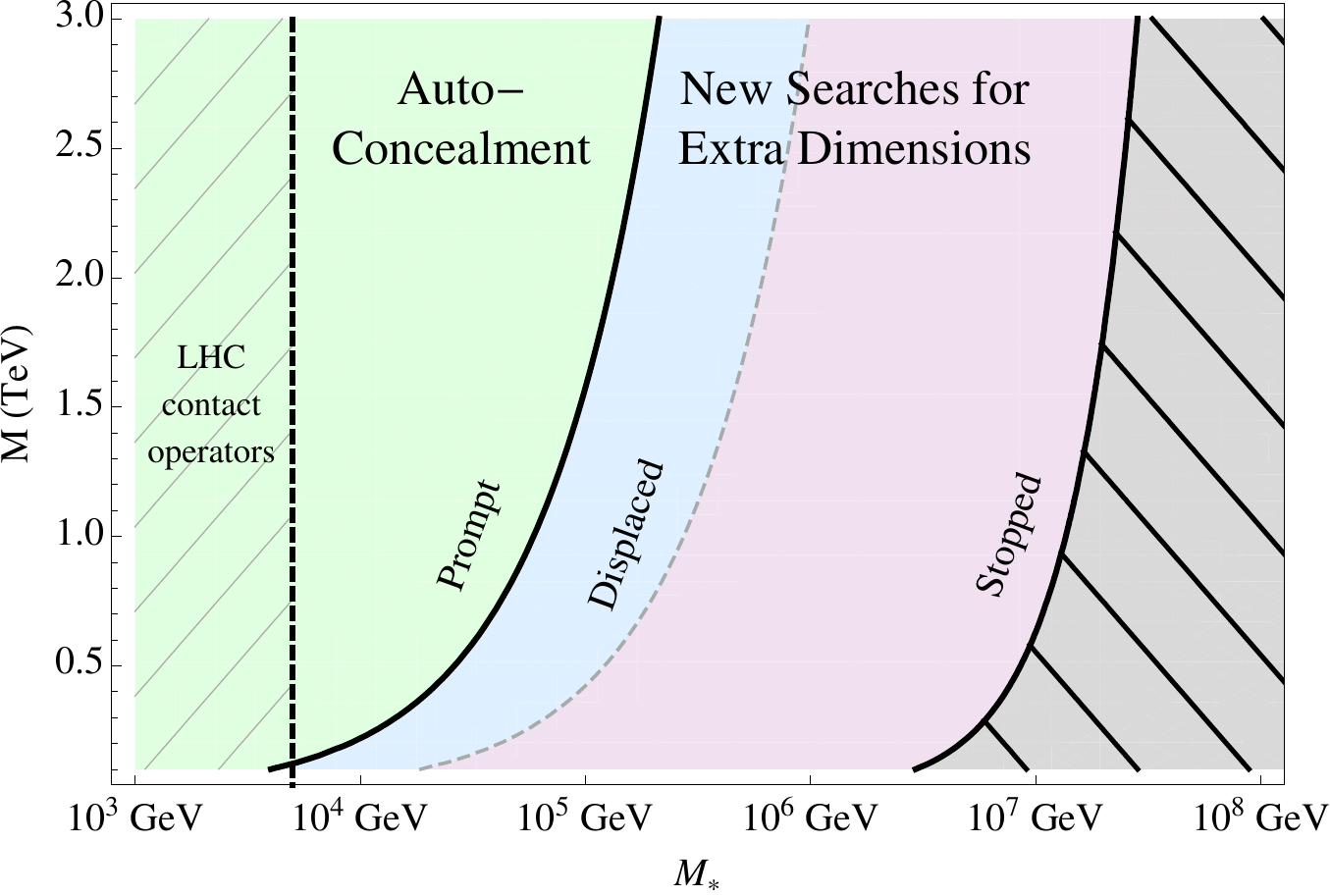}
  \caption{Colored regions display the form of LOSP decay as a function of the LOSP mass, $M$, and the fundamental gravitational scale, $M_*$.  The bulk LSP
  is taken to be a modulino, the LOSP to be a sfermion, and we show the case $d=4$. The auto-concealment mechanism applies in the region of prompt decays.  In the regions
  of displaced decays or stopped LOSP out-of-time decays the auto-concealment mechanism no longer functions, but the decays of the LOSP can  provide a new search mechanism for extra-dimensions with reach much greater than that provided by contact operators. In the grey hatched region to the far right, the splitting between KK states becomes large compared to the mass of the LOSP, $1/(M L) \gtrsim 0.1$ (all of the decays to the left of this region have lifetimes $\tau \lesssim 1{\rm~yr}$). The hatched region to the far left shows the range of $M_*$ excluded by current LHC contact operator searches for extra-dimensions.}
    \label{fig:SummaryPlot}
\end{figure}

If superpartners \emph{are} discovered at future colliders, then observations of the LOSP decay can be the leading signature of the extra-dimensional nature of the theory. Some probes of the properties of bulk states through prompt decays of a new non-supersymmetric colored states have been studied in refs.~\cite{Dienes:2012yz,Dienes:2014bka}. Because the visible sparticles are charged under the SM gauge group and brane-localized, their production and subsequent decay will be the dominant production mechanism for bulk modes, especially when $M_*$ is so large that the LOSP decay is displaced or occurs after the LOSP is stopped in a detector.  This can extend the reach for the fundamental gravitational scale as high as $M_* \lsim 10^9~\GeV$, far above the reach of the usual contact-operator based searches for extra dimensions.  As an example of the scales of interest, in Figure~\ref{fig:SummaryPlot} we show the relevant regions of the $M_*$-$M$ plane for the case of a bulk modulino LSP with $d=4$ extra dimensions (Figure~\ref{fig:SummaryPlotsd2&6} in Section~\ref{sec:probing} shows the $d=2,6$ cases).

\section{Decays to the bulk}

We now turn to a detailed discussion of the mechanism.   The decay of a brane-localised LOSP of mass $M$ into a bulk state propagating in $d$ extra dimensions of size  $L\gg1/M$ can be described by an effective theory for the bulk-brane interactions \cite{Mirabelli:1997aj,Sundrum:1998sj,ArkaniHamed:2001tb,Marti:2001iw,Hebecker:2001ke,Linch:2002wg,Hewett:2002uq}.   The description of the brane states as point-localized objects in the $d$ bulk dimensions is taken to be valid up to a scale $\Lambda_b < M_*$, where $M_*$ is the fundamental gravitational scale of the theory, and $M<\Lambda_b$ by assumption so that the decay is well described by the effective theory.\footnote{At distances shorter than $1/\Lambda_b$, the embedding of the brane in the $d$ bulk dimensions may be non-trivial; these effects could be taken into account by the presence of higher dimensional operators including terms with bulk derivatives. The scale $\Lambda_b$ could correspond to the fundamental gravitational scale $M_*$ or to an intermediate scale related to the extension of the brane embedding in the transverse directions.} 

To be concrete we start by studying the decays of a brane-localized $\tilde{e}_R$ LOSP to the fermion $\psi$ of a bulk chiral multiplet\footnote{The bulk theory has at least $N=2$ extended SUSY from the 4d perspective, and this N=1 `chiral multiplet' must in fact have bulk partners that fill out a full higher dimensional hyper-multiplet or vector multiplet, although these states need not couple to the brane. We use the N=1 superfield field notation of Ref.~\cite{ArkaniHamed:2001tb}.} $\Phi$, and then generalize to other interesting cases.  While we now focus on this case as a simple example, there are a variety of other strongly motivated possibilities. In addition to a slepton LOSP, the case of a stop/sbottom LOSP and the case of degenerate first and second generation squark LOSPs provide particularly interesting examples from the point of view of collider phenomenology which we study in detail in the following section.  The results derived in this section apply to any sfermion decaying to its massless SM fermion partner and a bulk modulino.  In Section~\ref{sec:varieties} we describe the form of the distribution for a general set of possible LOSPs and a variety of bulk LSP candidates.

\subsection{Bulk spectrum and profiles}

We study the bulk states by expanding in KK modes in the extra dimensions,
$$\psi = \sum_n \frac{1}{\sqrt{V}} f_n(y_i) \psi_n(x),$$
where $x$ are the (3+1) coordinates, $y_i$ are the extra bulk coordinates, $V$ is the volume of the bulk, and each KK mode has mass $m_n$. In flat extra dimensions and in the absence of any bulk mass terms for the state, there is a zero mode, $m_0 = 0$ and the splittings between KK modes are of order the size of the bulk $\Delta m_n \approx 1/L$. We will be interested in cases where the decays from the brane states are highly-localized compared to the size of the bulk; in this case, the decays are insensitive to the exact form of the boundary conditions for bulk fields far away from the MSSM brane and can be well described in the continuum approximation, $\Delta m_n \rightarrow 0$.

The spectrum of KK masses and profiles of a bulk multiplet will be perturbed by the presence of mass terms, which may be spread along the entire $4+d$ dimensional space occupied by the bulk state or be  localized in some of the extra dimensions (for example localized on the MSSM brane).  A mass term $m_{4+d}$ spread along the full ($4+d$) dimensional space lifts the start of the KK tower to $m_{4+d}$. We assume such terms are negligible compared to the mass scale of the decays, and will describe scenarios where this occurs in Section~\ref{sec:varieties}. Mass terms that are localized in some of the $d$ dimensions have their effects suppressed by the volume of the remaining space, and are generally only relevant if they are localized near the MSSM brane, in which case they can affect the wavefunction profiles near the brane $f_n(0)$.

For example, a mass term for the fermion components of $\Phi$ localized on the MSSM brane has the form
\beq
\mathcal{L} = \delta^d(y)\frac{( \mu \psi \psi)}{\Lambda^{d}_b} + h.c.
\label{eq:branemass}
\eeq
where the fermion $\psi$ is normalized as a bulk field with mass dimension $(3+d)/2$. 
The effect of the on-brane mass is to suppress the profile $f_n(0)$ of the KK states near the brane, which suppresses the coupling to brane-localized states. For KK masses $m_n \ll \Lambda_b$ and co-dimension $d\geq3$, the perturbation of the wave function at the brane $f_n(0)$ is independent of $m_n$: for small perturbations $\mu \lesssim \Lambda_b$, $f_n(0)$ is unsuppressed, while for large perturbations $\mu \gg \Lambda_b$, $f_n(0)\rightarrow0$ and the leading operators coupling brane fields to the bulk field will be those containing bulk derivatives $\sim \frac{\nabla_y \psi}{\Lambda_b}$ (this latter case is the correct description for instance when orbifold conditions in the fundamental theory force the wavefunction to vanish on the brane). For co-dimension $d=1$, $f_n(0)\sim \frac{m_n}{\mu}$ for $m_n \lesssim \mu$, and for $d=2$ there is a logarithmic dependence on $m_n$. Overall, the localized mass terms typically increases the efficiency of auto-concealment by decreasing the relative coupling of lighter KK modes to the MSSM brane states. As the sizes of the localized mass terms $\mu$ are only weakly constrained, to be conservative we assume they are negligible for the rest of this work.


\subsection{Brane couplings and decays}
\label{sec:couplingsdecays}

For a simple and well-motivated example, we take $\Phi$ to  couple to the MSSM states as a modulus in the Kahler potential with a gravitationally suppressed coupling
\beq
\label{eq:Lmodulino}
\mathcal{L}=\delta^d(y)\left.\frac{1}{2}\left[\frac{(\Phi + \Phi^*){e_R}^*e_R }{M_*^{(d+2)/2}}\right]\right|_{\theta^4}.
\eeq
\begin{figure}[t]
  \centering
  \includegraphics[width=.7\columnwidth]{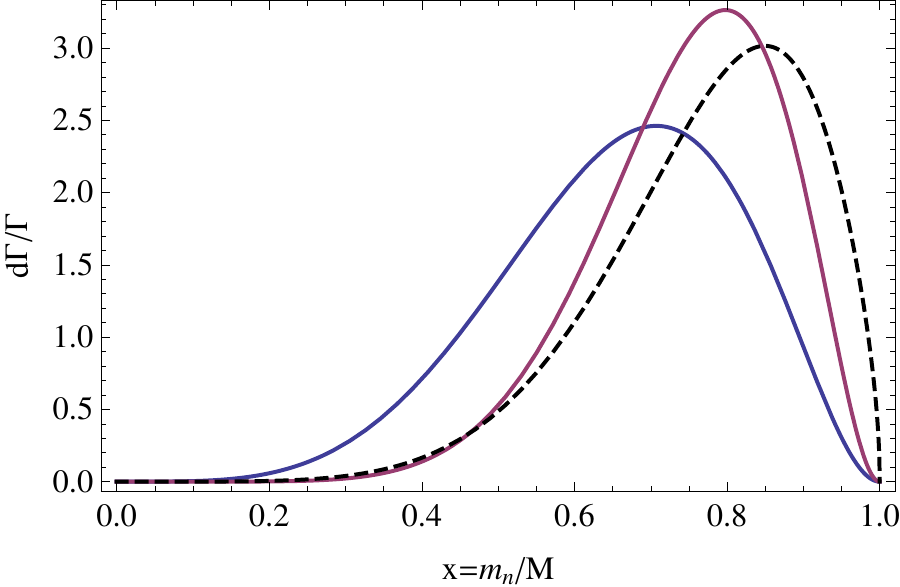}
  \caption{Differential distribution of KK masses for the decay $\tilde{e}_R\to e+\psi$ from Eq.(\ref{eq:SleptonModulino}) for $d=(3,6)$ (solid curves with peaks from left to right, respectively). Also shown dashed is the distribution for a $500$ GeV stop decaying in $d=6$ as $\tilde{t}_R\to t+\psi$, with the definition $x\equiv m_n/(m_{\tilde{t}_R}-m_t)$.}
    \label{fig:SleptonModulino}
\end{figure}
After making the KK expansion, the decay rate of a selectron with mass $M$ to each individual mode of mass $m_n < M$ that follows from
Eq.(\ref{eq:Lmodulino}) is
\beq
\label{eq:SleptonModulinoKK}
\Gamma_n = \frac{M^3}{8\pi M_*^{2+d}V}\frac{m_n^2}{M^2}\left(1-\frac{m_n^2}{M^2}\right)^2 .
\eeq
For co-dimension $d$, the number of states with mass $\sim m_n$ grows as $\sim m_n^{d-1}$ (this assumes the extra $d$-dimensions are flat---we later comment on the more general
case \cite{Kaloper:2000jb}). For this particular example, the rate to heavier KK states is further enhanced by a factor $m_n^2/M^2$ due to a helicity suppression of decays to lighter modes. Therefore even in $d=1$ the distribution will be peaked towards higher KK masses-- the extra-dimensional nature of the LSP is still crucial to provide the continuum of accessible states, but the enhancement of decays to heavier states is due completely to the matrix element. Going to the continuum limit, the total decay rate is
\beq
\label{eq:SleptonModulino}\Gamma_{\rm tot} = \sum_n^{m_n<M} \Gamma_n = \frac{M^{3+d}}{8\pi M_*^{2+d}} \frac{\Omega_{d}}{(2\pi)^{d}} \int_0^1 x^{d+1} (1-x^2)^2 dx,
\eeq
where $\Omega_d$ is the surface area of a $(d-1)$-sphere and $x\equiv m_n/M$. The resulting differential decay rate with respect to the KK mass of the modulino is shown in Figure~\ref{fig:SleptonModulino}. The most likely KK mass is $\sim(0.6\div 0.8) M$, and this can have striking observable consequences for collider phenomenology.  (For the case of a stop LOSP with decay $\tilde{t}\rightarrow t+\psi$, the non-negligible top mass modifies the distribution as shown in Figure~\ref{fig:SleptonModulino}.)

\section{SUSY limits and auto-concealment}
\label{sec:collider}

To understand the effect of auto-concealment on collider searches, it is useful to consider the limit that the LOSP decays to a very narrow distribution of bulk LSP KK states peaked at $m_n \approx M$.  In this case there is no visible energy from the LOSP decay\footnote{Decays of the bulk KK states among themselves producing visible energy on the brane are possible, but they are irrelevant on collider time scales due to the volume suppression of couplings to the brane.}, and events involving only direct pair production of the LOSP are invisible at colliders. This is identical to the case of exactly degenerate compressed spectra~\cite{Dreiner:2012gx}. In this kinematic limit, missing and visible transverse energy arise only when the system recoils against a radiated jet or photon--dominantly initial state radiation (ISR)--and SUSY searches are significantly weakened.  

A realistic distribution of KK masses as shown in Figure~\ref{fig:SleptonModulino} does not completely realize this limit; the distributions peak below $M$ and they have a non-negligible width. Nonetheless, they remain in the regime where most LOSP decays produce little visible energy and pair production events with large missing and visible energy are still dominantly due to hard ISR. The effect on experimental limits remains substantial. To illustrate this, we re-interpret existing 8 TeV LHC sparticle searches for three interesting cases of LOSP pair production followed by decays to a bulk modulino LSP: a right-handed slepton LOSP $\tilde{e}_R/\tilde{\mu}_R\to e/\mu+\psi$, a right-handed stop LOSP $\tilde{t}_R\to t+\psi$, and degenerate first and second generation squarks $\tilde{q}_{u,d,c,s}\to q+\psi$. We simulate sfermion pair production processes with MadGraph5 \cite{Alwall:2011uj} with shower and decays\footnote{To implement LOSP decays to a KK tower of fermion LSPs we introduced $N\sim 20$ new gauge neutral spin $1/2$ states in Pythia. The masses of these states $m_j$ fell into $N$ evenly spaced bins from 0 to the LOSP mass $M$.  The mass $m_j$ of the $j^{\rm{th}}$ state was given by the branching ratio-weighted average of masses in the $j^{\rm{th}}$ bin, and the branching fraction to this state was determined by the integrated width over the bin.} in Pythia6 \cite{Sjostrand:2006za} and MLM matching of up to one additional jet. With one exception,\footnote{With the exception of \cite{Aad:2014qaa}, all of the analysis used in this paper to recast limits have been validated by CheckMATE.  It was felt important to include this unvalidated analysis since it provided the only exclusion limits for stops decaying to a modulino in $d=6$.} experimental limits were recast using validated analyses in CheckMATE \cite{Drees:2013wra,deFavereau:2013fsa,Cacciari:2005hq,Cacciari:2011ma,Cacciari:2008gp,Read:2002hq}.  While we expect the results of these simulations to broadly characterize how auto-concealment affects current supersymmetry search limits, it is up to the experimental collaborations to set definitive bounds.

The first process we consider is pair production of degenerate right-handed sleptons decaying to a bulk modulino $\tilde{e}_R/\tilde{\mu}_R\to e/\mu+\psi$. 
The dominant limit is from a $20.3~\rm{fb}^{-1}$ ATLAS $l^+ l^- + \MET$ search \cite{ATLAS-CONF-2013-049} based on the kinematic variable $m_{T2}$ \cite{Lester:1999tx,Barr:2003rg,Cheng:2008hk}. The effect of auto-concealment on missing energy-related observables is dramatic, as illustrated in Figure~\ref{fig:MT2}, which shows the signal $m_{T2}$ distribution after typical cuts used to reduce backgrounds.  For the case of $d=3$, the number of events satisfying the signal region cuts is very significantly reduced, while for $d=6$ essentially no events pass cuts for the illustrated case of $M_{{\tilde l}_R}=150~\GeV$ and $20.3\ifb$.  The effect on exclusion limits is predictable.  Figure~\ref{fig:sleptonxsplot} shows the strongest cross section exclusion limit (at 95\% $\rm CL_S$) from the ATLAS searches \cite{ATLAS-CONF-2013-049,ATLAS-CONF-2013-089}. A monojet search \cite{ATLAS-CONF-2012-147} was also considered to pick up ISR but the analysis had no effect on limits as it vetoed events with isolated leptons. The existing LHC8 limits of $M_{{\tilde l}_R}\gsim 225~\GeV$ for direct production of right handed sleptons decaying to a massless LSP are completely eliminated, with only the much weaker LEPII limit of $M_{{\tilde l}_R}\gsim 95~\GeV$ for very compressed slepton decays still applying \cite{LEPSUSYWG-04-01-1,Heister:2001nk,Achard:2003ge,Abbiendi:2003ji,Abdallah:2003xe}.\footnote{Note that direct production of left-handed sleptons is already concealed independent of the existence of a bulk LSP as the EW symmetry breaking mass splitting between the heavier charged and lighter neutral members of the LH slepton doublet is small enough that a compressed spectrum is automatically realised.}

\begin{figure}[h]
  \centering
  \includegraphics[width=.7\columnwidth]{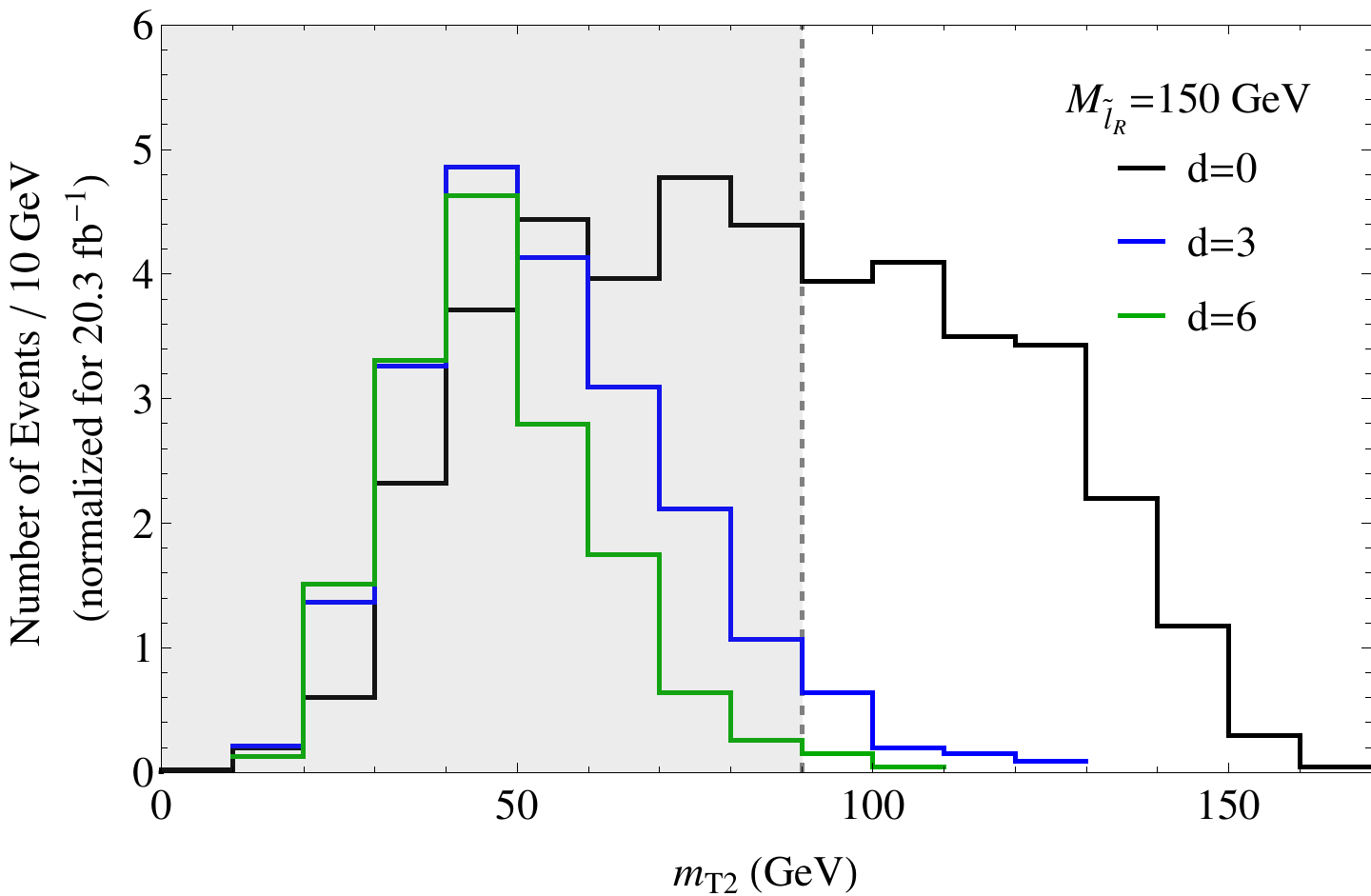}
  \caption{Differential distribution in stranverse mass $m_{T2}$ for the decay $\tilde{e}_R \rightarrow e + \psi$ for a slepton of mass $M=150~\gev$ to a single massless LSP (black) and a bulk modulino LSP (blue and green) as in Eq.(\ref{eq:SleptonModulino}) for $d=3,6$. The preselections of Ref.~\cite{ATLAS-CONF-2013-049} have been applied, including a cut on missing energy, $E_{T}^{\rm miss,rel}>40~\GeV$, which leads to the different total number of events for each case. Shown by a dashed line is the signal region cut $m_{T2}>90~\GeV$ used to reduce backgrounds such as $W^+W^-$ production.  Definitions of $E_{T}^{\rm miss,rel}$ and $m_{T2}$ can be found within Ref.~\cite{ATLAS-CONF-2013-049}.}
    \label{fig:MT2}
\end{figure}

\begin{figure}[h]
  \centering
  \includegraphics[width=0.7\columnwidth]{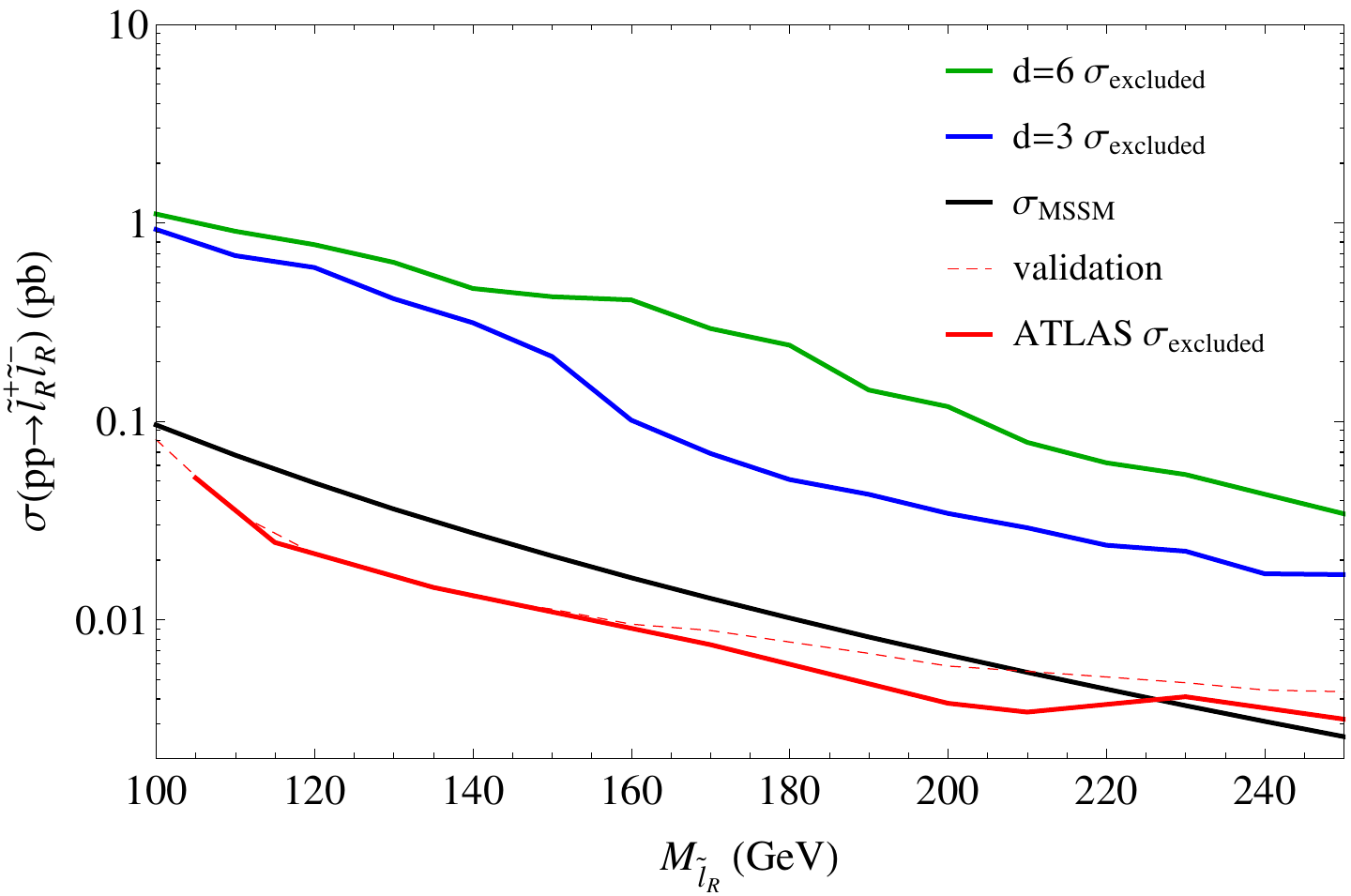}
  \caption{Strongest upper bound on degenerate ${\tilde \mu_R}, {\tilde e_R}$ slepton pair production cross sections from ATLAS $l^+ l^- + \MET$  $m_{T2}$ \cite{ATLAS-CONF-2013-049} and razor analyses \cite{ATLAS-CONF-2013-089}.  A monojet search was also considered \cite{ATLAS-CONF-2012-147} but did not affect limits. The top two curves corresponds to sleptons promptly decaying to the KK tower of a massless modulino in $d=3$ (blue) and $d=6$ (green) extra dimensions.  The $m_{T2}$ analysis is more effective at higher masses; below 140 GeV (170 GeV) for $d=3$ ($d=6$) the razor analysis sets stronger limits.  Solid red (lowest) curve gives the observed ATLAS upper bound on the RH slepton production cross section from \cite{ATLAS-CONF-2013-049} for decays to a massless LSP. For validation, a dashed red curve gives the same bound using our simulation. Black curve gives the predicted NLO direct production cross section~\cite{Beenakker:1999xh} with other superpartners decoupled, illustrating that RH sleptons are excluded up to $\sim 225~\GeV$ for decays to a massless LSP.  For the searches considered, present limits on direct production of RH sleptons evaporate in the presence of the auto-concealment mechanism.}
    \label{fig:sleptonxsplot}
\end{figure}

Limits on 3rd generation squark production can also be dramatically reduced. We studied $\tilde{t}_R$ pair production with $\tilde{t}_R \to t+\psi$. As depicted in Figure~\ref{fig:SleptonModulino}, the distribution of KK states in the decay is slightly modified from the result for a massless SM fermion  fermion, Eq.(\ref{eq:SleptonModulino}),  due to the non-negligible top mass. The dominant validated analysis in CheckMATE was the ATLAS $20.3~\rm{fb}^{-1}$ all hadronic $6~ (2~b){\rm~jet~} + \MET$ search~\cite{ATLAS-CONF-2013-024}, while the unvalidated 2 lepton stop search \cite{Aad:2014qaa} provided the strongest limits below $\sim360~\GeV$.  Figure \ref{fig:stopxsplot} shows cross section limits for
these searches. For prompt decays to a massless LSP the limit is $m_{\tilde{t}}\gtrsim680\GeV$, while limits reduce to $\sim350\div410~\GeV$ for decays to a bulk modulino in $d=3,6$. A number of other searches are expected to provide similar limits, for example the ATLAS and CMS  semi-leptonic searches~\cite{CMS-PAS-SUS-14-011,Aad:2014kra} and the most recent all-hadronic searches~\cite{CMS-PAS-SUS-14-011,Aad:2014bva} which perform better than \cite{ATLAS-CONF-2013-024} at low stop masses in the compressed region.

\begin{figure}[h]
  \centering
  \includegraphics[width=0.7\columnwidth]{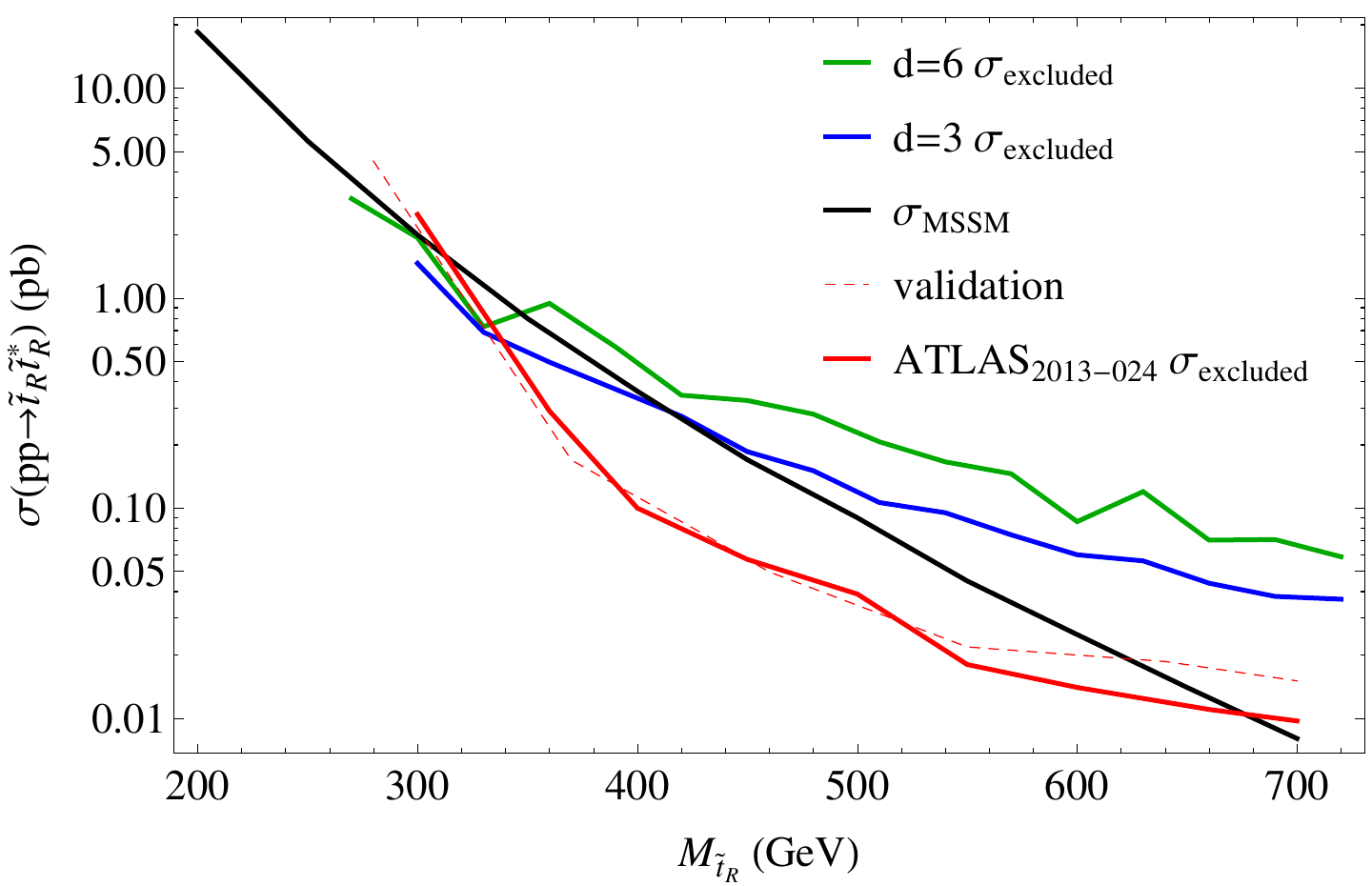}
  \caption{Strongest upper bound on stop pair production cross sections from ATLAS $6~ (2~b){\rm~jet} + \MET$  \cite{ATLAS-CONF-2013-024} and 2 lepton stop \cite{Aad:2014qaa} searches.  A razor analysis \cite{ATLAS-CONF-2013-089}, a one lepton stop search \cite{Aad:2014kra}, and two monojet searches \cite{ATLAS-CONF-2012-147,Aad:2014nra} were also considered but did not strengthen the exclusion limits.  The upper two curves corresponds to stops promptly decaying to a top + the KK tower of a massless modulino in $d=3$ (blue) and $d=6$ (green) extra dimensions.   The all hadronic analysis is more effective at higher masses; below $\sim360$ GeV the two lepton analysis sets stronger limits, however it should be noted that this analysis is not yet validated by CheckMATE.  Solid red (lowest) curve gives the observed ATLAS upper bound on the stop production cross section from \cite{ATLAS-CONF-2013-024} assuming prompt decay to a top + a massless LSP. For validation, a dashed red curve gives the same bound using our simulation.  Black curve gives the predicted NLO direct production cross section~\cite{LHCSUSYXSWG,Kramer:2012bx}, thus illustrating that stops are excluded up to $\sim 680~\GeV$ for a single massless LSP.  For the search considered, present limits on direct production of stops drop to $\sim 350\div410~\GeV$ in the presence of the auto-concealment mechanism.} 
    \label{fig:stopxsplot}
\end{figure}

We finally study pair production of degenerate first and second generation squarks with $\tilde{q}_i \to q_i+\psi$ assuming the gluinos and 3rd generation squarks are decoupled. The dominant limits shown in Figure~\ref{fig:squarkxsplot} are from the ATLAS $20.3~\rm{fb}^{-1}$ ${2-6\rm~jets~} + \MET$ analysis~\cite{ATLAS-CONF-2013-047}, except for squarks below 200 GeV where limits are driven by the monojet search \cite{ATLAS-CONF-2012-147}.  These searches have hard cuts on missing and visible energy and are substantially affected by auto-concealment. While for a decay to a single massless LSP the limit is $M\gtrsim 800~\GeV$, for decays to a bulk modulino in $d=3,6$ the limit is reduced to only $\sim450~\GeV$.  We have assumed no D-term splitting leading to decays between the left handed squarks, but we do not expect that these soft decays would significantly affect the results.

\begin{figure}[h]
  \centering
  \includegraphics[width=0.7\columnwidth]{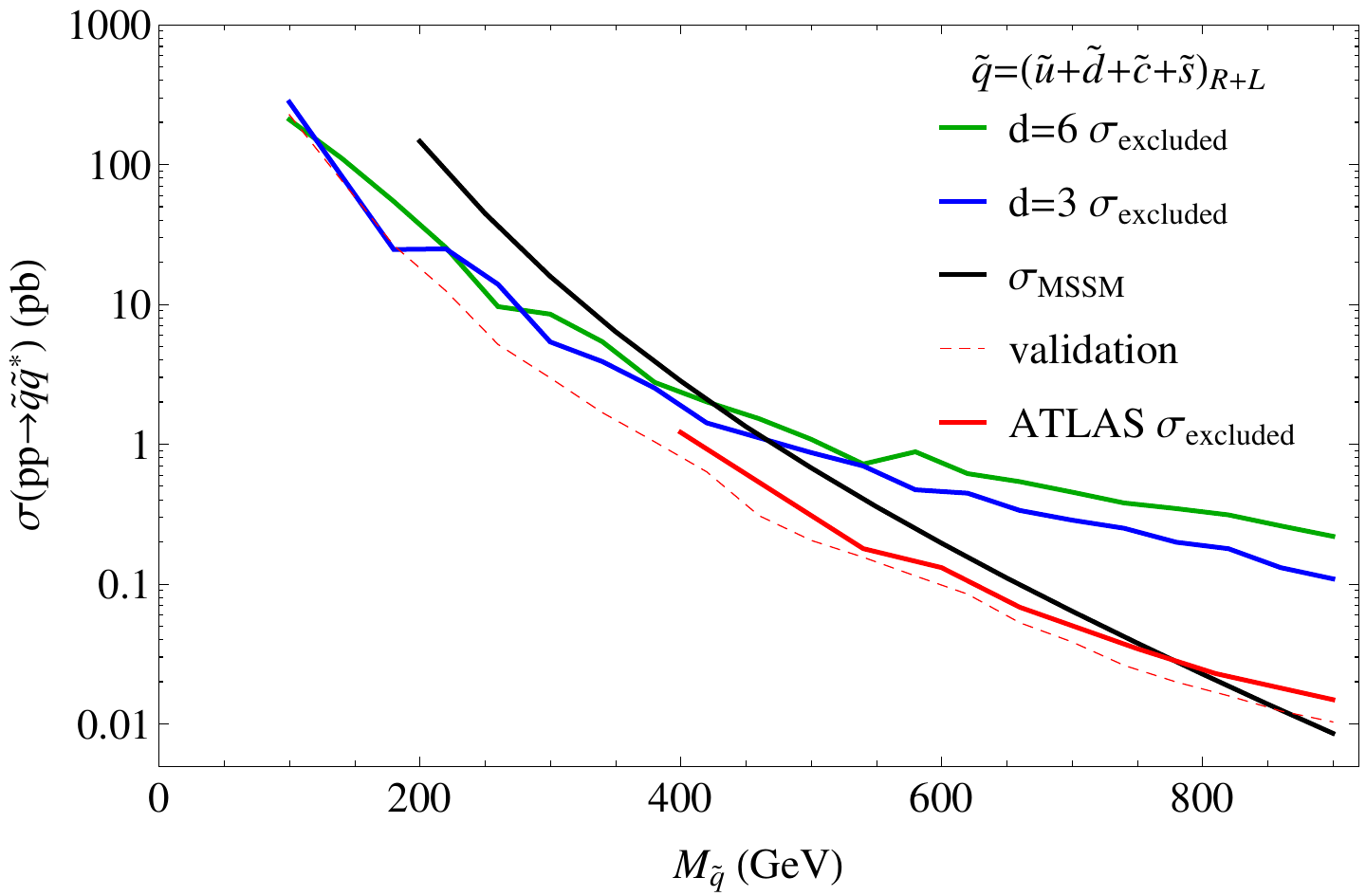}
  \caption{
    Strongest upper bound on pair production cross sections for degenerate first and second generation squarks from ATLAS ${2-6\rm~jets} + \MET$ \cite{ATLAS-CONF-2013-047} and monojet \cite{ATLAS-CONF-2012-147} searches.  A razor analysis was also considered \cite{ATLAS-CONF-2013-089} but its limits were weaker.  The top two curves corresponds to squarks promptly decaying to the KK tower of a modulino in $d=3$ (blue) and $d=6$ (green) extra dimensions.  The hadronic search is the more effective of the two analysis except below $\sim200$ GeV.  Solid red (lowest) curve gives the observed ATLAS upper bound on the squark production cross section from \cite{ATLAS-CONF-2013-047} assuming prompt decay to a LSP with mass $\sim40~\rm{GeV}$. Dashed red curve gives our bounds for a single massless LSP for validation. Black curve gives the predicted NLO direct production cross section when gluinos are decoupled~\cite{LHCSUSYXSWG,Kramer:2012bx}, thus illustrating that degenerate squarks are excluded up to $\sim 775~\GeV$ for a single massless LSP.  For the searches considered, present limits on direct production of squarks drops to $\sim 450~\GeV$ for $d=3,6$ in the presence of the auto-concealment mechanism.}
    \label{fig:squarkxsplot}
\end{figure}

\begin{figure}[h]
  \begin{minipage}{0.48\linewidth}
    \centering
    \includegraphics[scale=0.2]{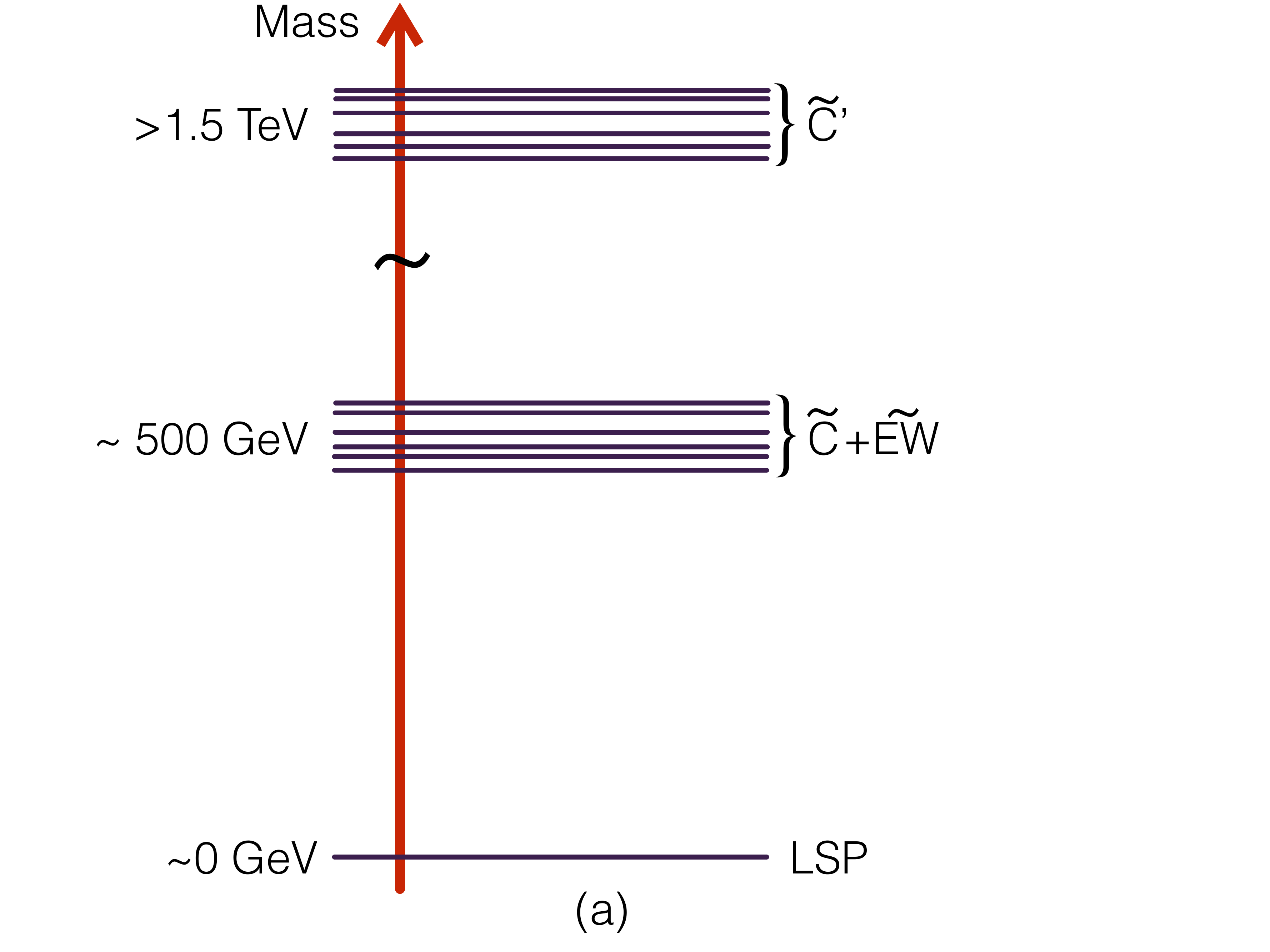}
  \end{minipage}
  \hspace{0.4 cm}
  \begin{minipage}{0.48\linewidth}
    \centering
    \includegraphics[scale=0.2]{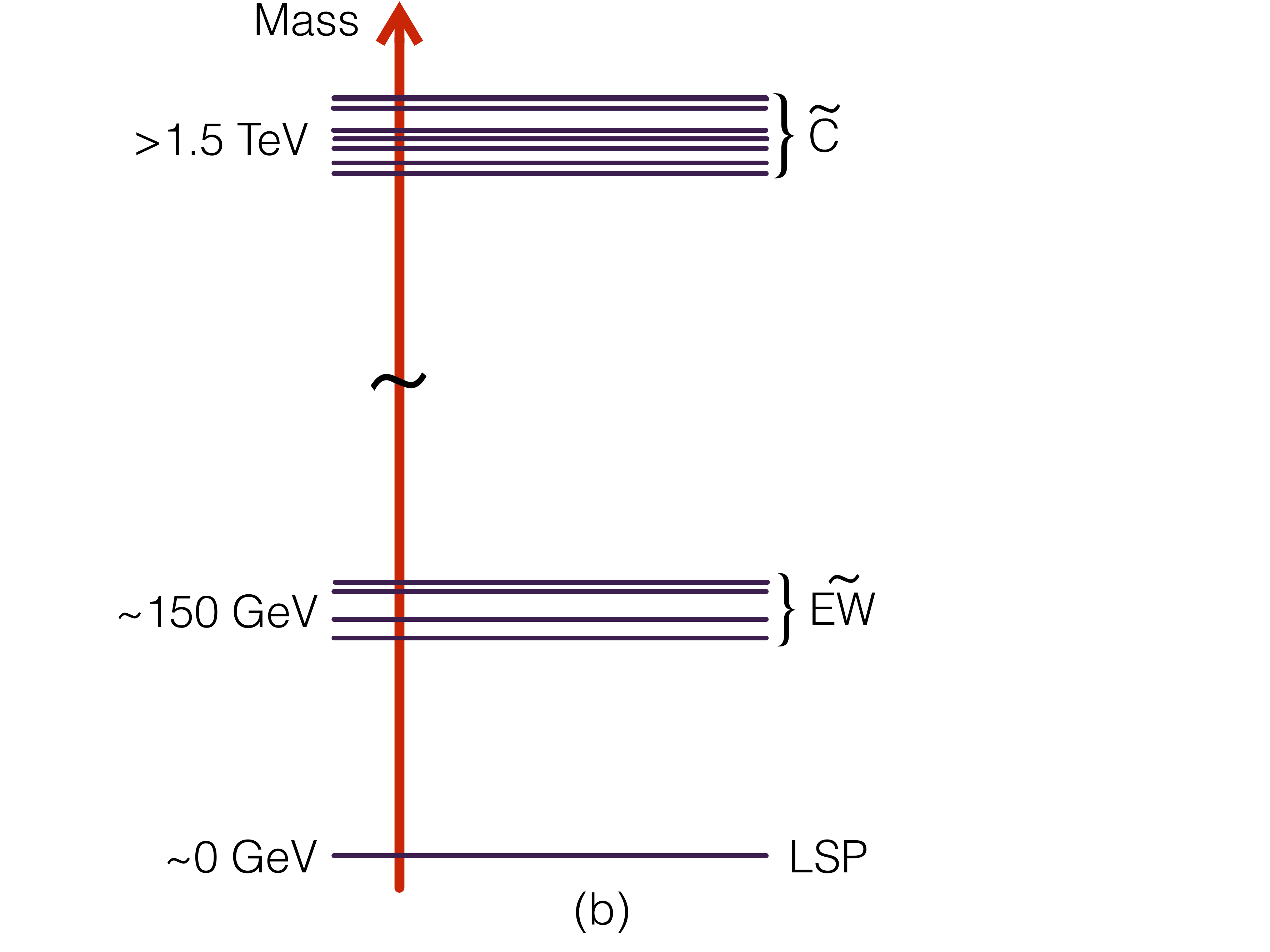}
  \end{minipage}\\
   \vspace{1.0 cm}\\
   \begin{minipage}{0.48\linewidth}
    \centering
    \includegraphics[scale=0.2]{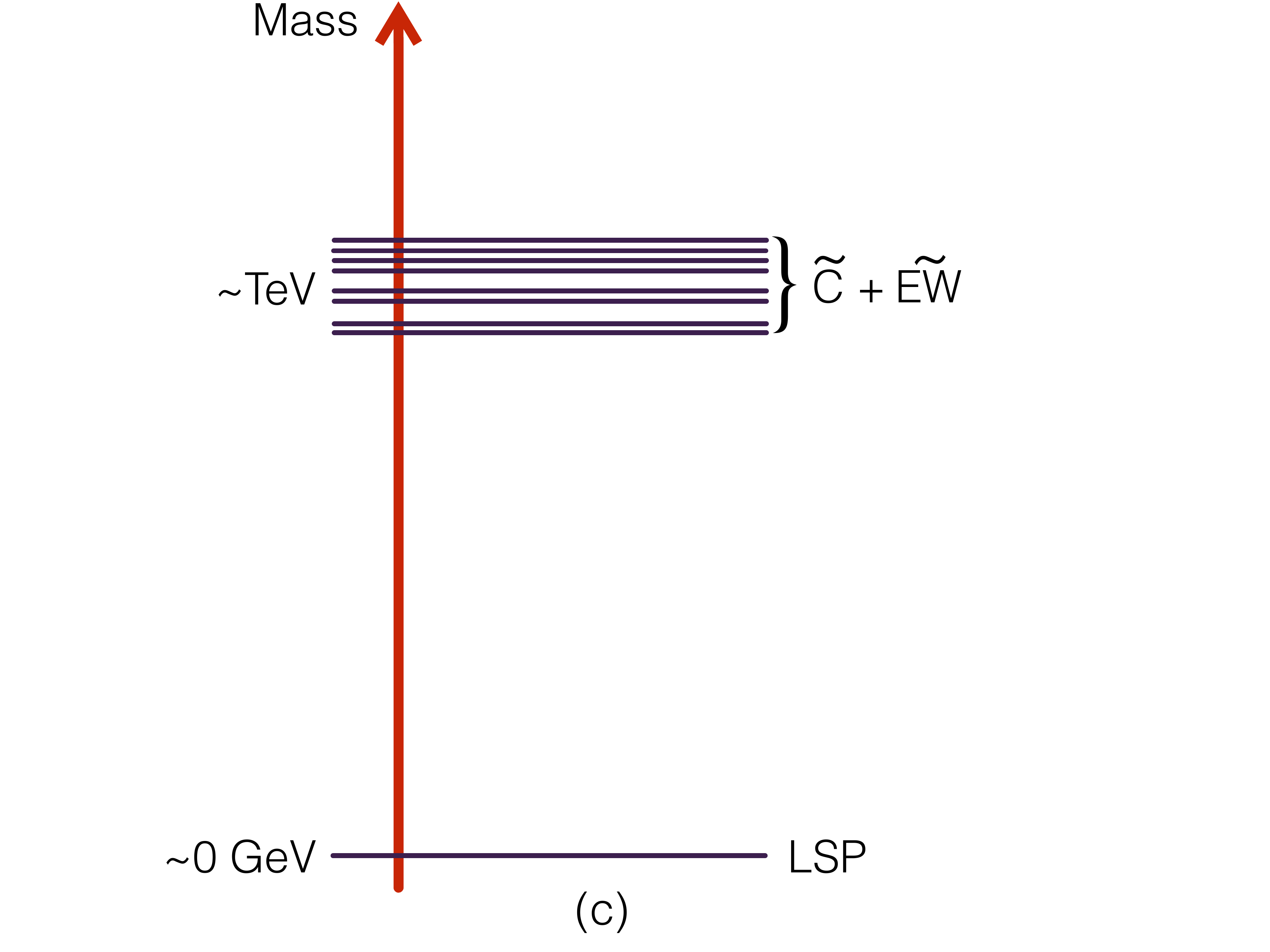}
  \end{minipage}
  \hspace{0.4 cm}
  \begin{minipage}{0.48\linewidth}
    \centering
    \includegraphics[scale=0.2]{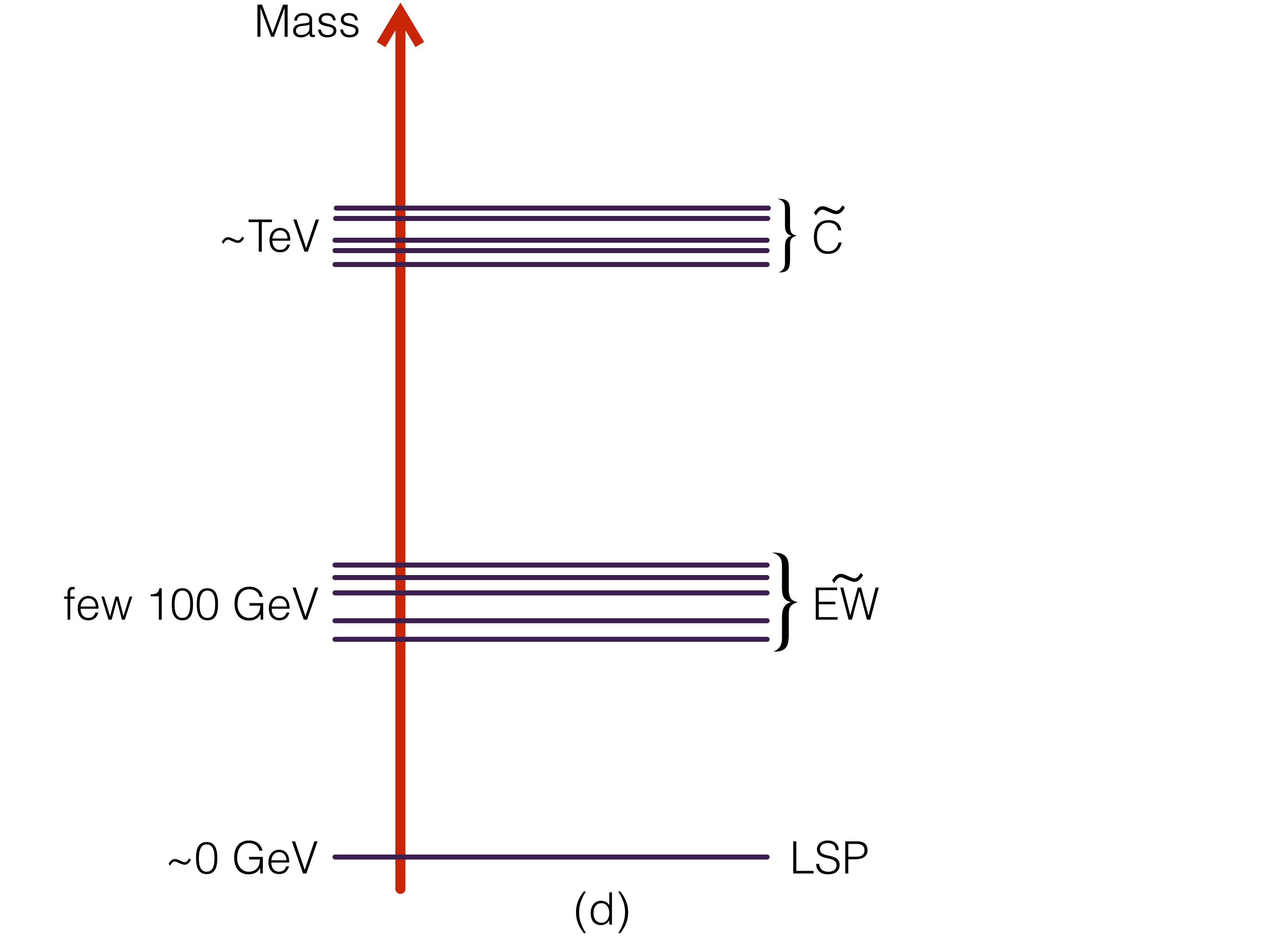}
  \end{minipage}
\caption{Schematic sparticle spectra for which efficient auto-concealment applies, cases (a)-(c), and for which it fails (d).  In case
(d) colored particles, here generically denoted $\tilde C$, are light enough, $\sim 1~\TeV$, that they can be copiously pair produced at the LHC and simultaneously EW sparticles,
among which is the LOSP, are substantially less massive, $\sim {\rm few}\times 10^2~\GeV$.  In this situation there are many deep cascade decays of the heavier colored
particles to the LOSP producing both highly energetic jets, and a boosted LOSP which leads to a boost of the $\MET$ produced in the final decay of the LOSP.  Conversely, in
case (a) all the heavier colored sparticles, $\tilde C'$ are so massive as to have small pair production cross sections at the LHC, while the lighter colored sparticles (\eg, the 3rd
generation squarks) are not too far separated from the EW sparticles, so only leading to shallow cascade decays before the final decay of the LOSP to the LSP KK tower.
In (b) all colored particles are too heavy to be substantially produced at the LHC $m_{\tilde C}\gsim 1.5~\TeV$, while the copiously produced light EW sparticles have only shallow cascade decays before LOSP decay. In case (c) all sparticles apart from the LSP are moderately heavy and roughly comparable in mass, and the dominant production through the colored states undergoes only shallow cascades to the LOSP.}
\label{fig:sparticlespectra}
\end{figure}

We have seen that auto-concealment significantly reduces bounds on direct production of superpartners,  dynamically realizing the signatures of a compressed spectrum where a single LSP is nearly degenerate with the LOSP. It is important to emphasise that the auto-concealment mechanism, like the compressed case, does {\it not} alleviate bounds on all forms of visible sparticle spectra. This is because of the possibility of highly energetic cascade decays before the decay of the LOSP.  To distinguish the bad cases from the good it is useful to define {\it deep} cascade decays as ones where the splitting between the parent visible-sector sparticle and the LOSP are large, $\Delta {\tilde m} \gtrsim M$, and conversely, {\it shallow} cascade decays as
ones involving parent-LOSP splittings $\Delta {\tilde m} \lesssim M$.   Auto-concealment does not substantially ease bounds on spectra driven by large cross-sections
for deep cascade decays to the LOSP.   The reason for the failure of efficient auto-concealment in this case
is that deep cascades produce highly energetic visible particles (\eg, jets or leptons)  recoiling from a highly boosted LOSP, which is transformed primarily into $\MET$ in the final decay of the LOSP to the LSP KK tower. 

A common example where auto-concealment fails to improve limits is when squark LOSPs of mass $M$ are accompanied by a gluino of mass $\lesssim 2M$ \cite{Arvanitaki:2013yja}; even though direct production limits could allow squark LOSPs as light as $\sim450\GeV$, we estimate gluino pair and associated production with decays to squarks sets much stronger limits $M_g \gsim (1.5\div 2)~\TeV \gg  2\times(450~\GeV)$. On the other hand there are a variety of visible-sector sparticle spectra for which auto-concealment is efficient. For example Dirac gluinos can naturally have a mass $M_g \gg 2M$, yielding a sufficiently small production cross section for deep cascades that auto-concealment is effective. To illustrate this point more generally we show in Figure~\ref{fig:sparticlespectra}
four examples of visible-sector spectra, one of which (d) fails the condition for efficient auto-concealment, while (a)-(c) satisfy the condition.  

A number of search strategies using leptons to detect compressed SUSY spectra have been developed (since leptons typically have softer $p_T$ cuts than jets) and will likely be useful in detecting shallow decays and discovering auto-concealment signatures.  For this purpose ref.~\cite{Buckley:2013kua} developed a $ll+\MET$ ``super-razor" search, while ref.~\cite{Rolbiecki:2012gn} proposed modifying cuts on existing lepton searches.  Additionally, refs.~\cite{Giudice:2010wb,Schwaller:2013baa, Baer:2014kya} have suggested modifying monojet searches to include soft leptons as a way of picking out shallow electroweak decays with hard ISR.

\section{Probing extra dimensions}
\label{sec:probing}

While decays of the LOSP to a KK tower of a bulk LSP can erode limits from standard promptly decaying super-partner searches, they also have the potential to open a new window for probing extra dimensions if SUSY particles are discovered. Decays of the LOSP are potentially observable over a wide range of lifetimes through prompt decays in the detector, in-flight decays ($1{\rm~mm}\lesssim c\tau \lesssim 10{\rm~m}$), or decays of stopped particles ($100{\rm~ns} \lesssim\tau \lesssim 1{\rm~yr}$). As shown in  Figures~\ref{fig:SummaryPlot} ($d=4$) and \ref{fig:SummaryPlotsd2&6} ($d=2,6$), this corresponds to range of fundamental scales that can far exceed the current reach\footnote{We assume the contact operators have $\mathcal{O}(1)$ coefficients suppressed by the scale $M_*$, corresponding to a weakly coupled UV completion of gravity.} $M_*\gtrsim5~\TeV$ \cite{Aad:2014wca,ATLAS:2012ky} of traditional collider searches for KK graviton emission and contact operators\footnote{In addition to contact operators for SM states, operators contributing to sfermion production have also been studied in refs. \cite{Hewett:2002uq, Baek:2002np}. The lower dimension contact operators contributing to sfermion production described in ref.~\cite{Hewett:2002uq} can set stronger bounds for certain SUSY spectra, but can also be forbidden if some R-symmetries are preserved near the brane to energies below the fundamental scale.}~\cite{Giudice:1998ck,Hewett:1998sn,Mirabelli:1998rt,Han:1998sg,Contino:2001nj,Giudice:2003tu,Giudice:2004mg} and searches for the effects of Higgs mixing with bulk states~\cite{Diener:2013xpa}.  This reach can also greatly exceed that of astrophysical searches\footnote{Astrophysical searches can also easily be avoided by modifying the low energy KK spectrum \cite{Kaloper:2000jb,Giudice:2004mg}.} \cite{Hannestad:2003yd} and reaches values of $M_*$ which are consistent with cosmological limits for a large range of reheat temperatures.
This motivates studying strategies to  distinguish decays to a bulk LSP from other scenarios, for instance two- and three-body decays to a single massive LSP, and further to distinguish different numbers of bulk dimensions and different bulk LSP candidates. A detailed study of this possibility is beyond the scope of this work, but we describe briefly some of the relevant issues.


\begin{figure}[t]
 \begin{minipage}{0.48\linewidth}
  \centering
  \includegraphics[width=.9\columnwidth]{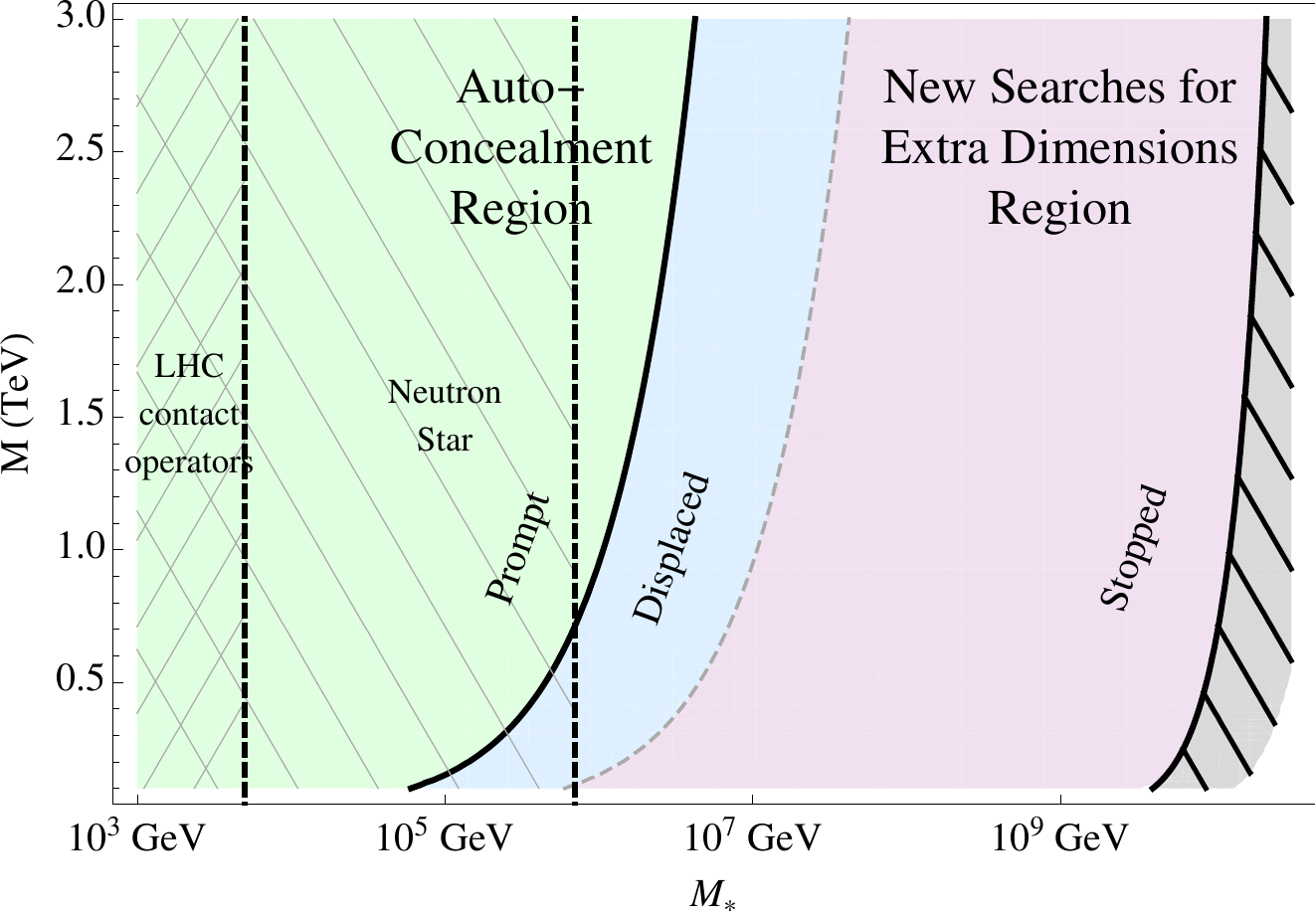}
   \end{minipage}
  \hspace{0.4 cm}
  \begin{minipage}{0.48\linewidth}
 \includegraphics[width=.9\columnwidth]{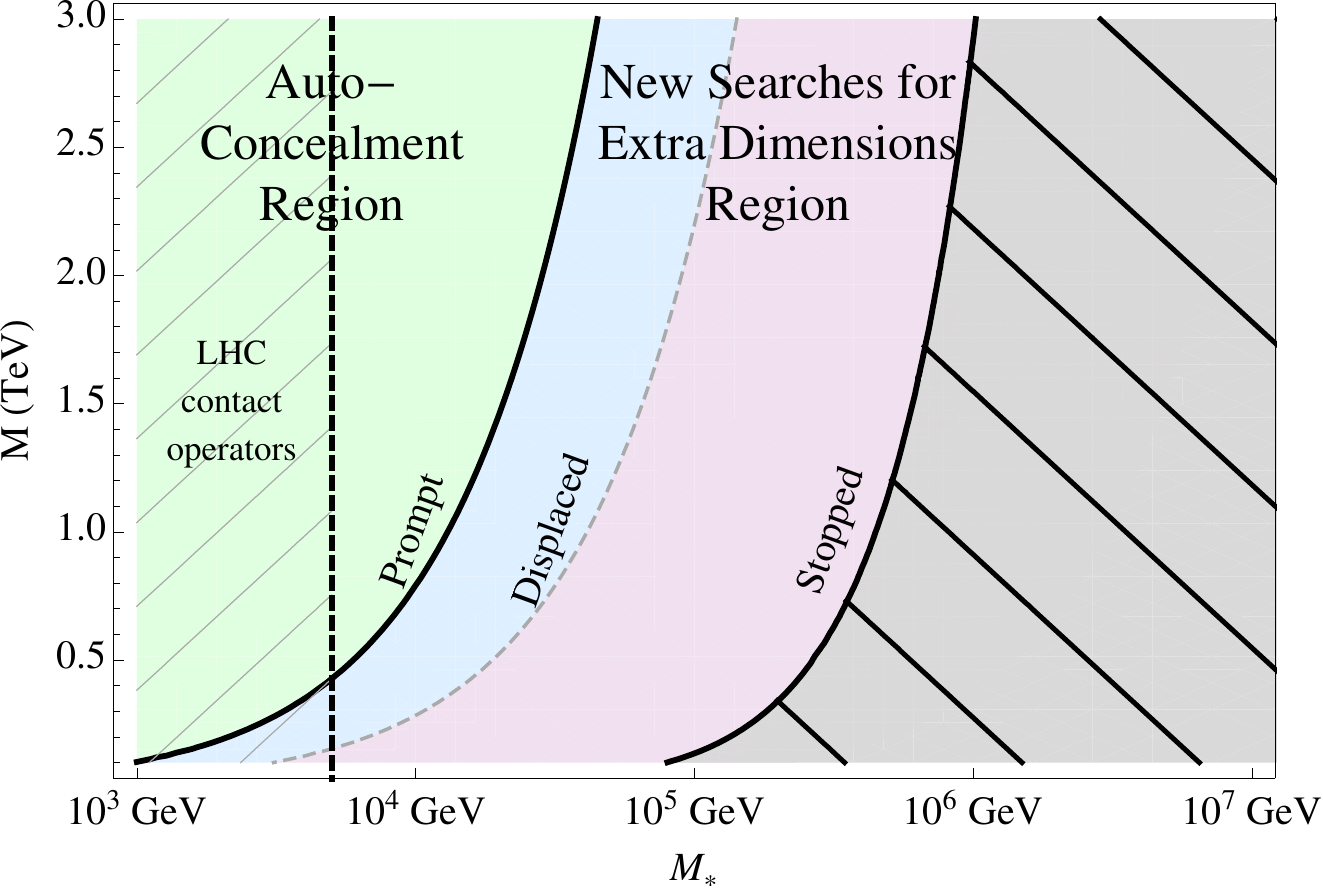}
 \end{minipage}
  \caption{Companion to Figure~\ref{fig:SummaryPlot} in the Introduction. Colored regions display the form of LOSP decay as a function of the LOSP mass, $M$, and the fundamental gravitational scale, $M_*$, for $d=2$ (left panel) and $d=6$ (right panel).  In the grey hatched region to the far right, the splitting between KK states becomes large compared to the mass of the LOSP, $1/(M L) \gtrsim 0.1$ (all of the decays to the left of this region have lifetimes $\tau \lesssim 1{\rm~yr}$). The hatched region to the far left shows the range of $M_*$ excluded by current LHC contact operator searches for extra-dimensions. In the left panel we also show the region that can potentially be excluded by neutron star observations~\cite{Hannestad:2003yd}.  This exclusion, however, depends
on the far IR part, $m_{KK}\lsim 100~\MeV$, of the KK spectrum of the graviton which is highly dependent on the assumption of perfectly flat extra dimensions.  In the more general case
of curved (but still unwarped) extra dimensions the neutron star limits no longer apply \cite{Kaloper:2000jb,Giudice:2004mg}.}
    \label{fig:SummaryPlotsd2&6}
\end{figure}

For prompt decays, a variety of LHC studies have demonstrated that features of $m_{T2}$ and similar generalized distributions could determine the LSP mass in some spectra (for a review, see ref.~\cite{Barr:2010zj}) using $100\div 1000$ BSM events after cuts (corresponding to $\sim2.0\div 2.5~\TeV$ squarks and gluinos at the high luminosity LHC13). Such observations should be sensitive to the absence of a single mass for the LSP, and may be adaptable to identify a bulk LSP. The possibility of measuring the properties of a distribution of dark matter particles with different masses at the LHC has been studied in the particular case of three-body decays of a scalar color octet in Refs.\cite{Dienes:2012yz,Dienes:2014bka}. For a light enough or electroweak dominated spectrum, measurements at a $e^+e^-$ collider may be more promising~\cite{MoortgatPick:2008yt,Djouadi:2007ik,Martyn:2004jc,Martyn:2004ew}.

If the LOSP decay is displaced in the detector or is long-lived on collider time scales then a great deal more information becomes accessible.  The lifetime can be directly measured \cite{Asai:2009ka,Pinfold:2010aq,Ito:2011xs} and if the LOSP is charged or colored, its mass is directly observable through timing and ionization measurements \cite{Hinchliffe:1998ys,Kilian:2004uj,Allanach:2001sd,Rajaraman:2007ae}. Proposals have been made to study the kinematics of production and in-flight decays \cite{Kitano:2008sa,Rajaraman:2007ae,Feng:2009yq,Luty:2011hi,Chang:2011jk} as well as decays of charged/colored LOSPs stopped in the existing LHC detectors \cite{Buchmuller:2004tm,Buchmuller:2004rq,Pinfold:2010aq,
Asai:2009ka,Ito:2011xs,Graham:2011ah} or a dedicated stopper-detector \cite{Hamaguchi:2004df,Feng:2004yi}. Tracking of particles from in-flight decays with large displacements or decays of LOSPs stopped in dense regions of detector material may be challenging. However, measurements of the kinematic features of in-flight and stopped decays have two primary advantages compared to techniques for prompt decays:
i) backgrounds are very small and require fewer cuts which affect the kinematic distributions of the decay, 
ii) the rest frame and mass of the parent particle can be determined independently from the decay. 
Combining sufficiently precise measurements of the LOSP decay with mass and lifetime measurements could  give strong evidence for the nature of the bulk LSP, the number of bulk dimensions, and the scale $M_*$!

\section{Varieties of bulk LSPs}
\label{sec:varieties}

We are interested in models where the MSSM particles are confined to a brane in a gravitational bulk with $d$ additional compact dimensions of size $L \gg \TeV^{-1}$.  At distances $\ll \TeV^{-1}$, the bulk and the MSSM brane are locally supersymmetric, with at least an $N=1$ subset of the bulk supersymmetries realized on the MSSM brane. While the MSSM brane must be localized within the large compact dimensions, at distances $\lesssim \TeV^{-1}$ some or all of the MSSM states may extend around additional small dimensions or cycles, and other branes of various dimensions may also be present. This set-up is illustrated in Fig.~\ref{fig:SUSYBreaking}(a), and allows the realization of a variety of extra-dimensional SUSY breaking mechanisms, sequestered sectors, and string embeddings of the MSSM structure. 

\begin{figure}[t]
  \centering
  \includegraphics[width=.9\columnwidth]{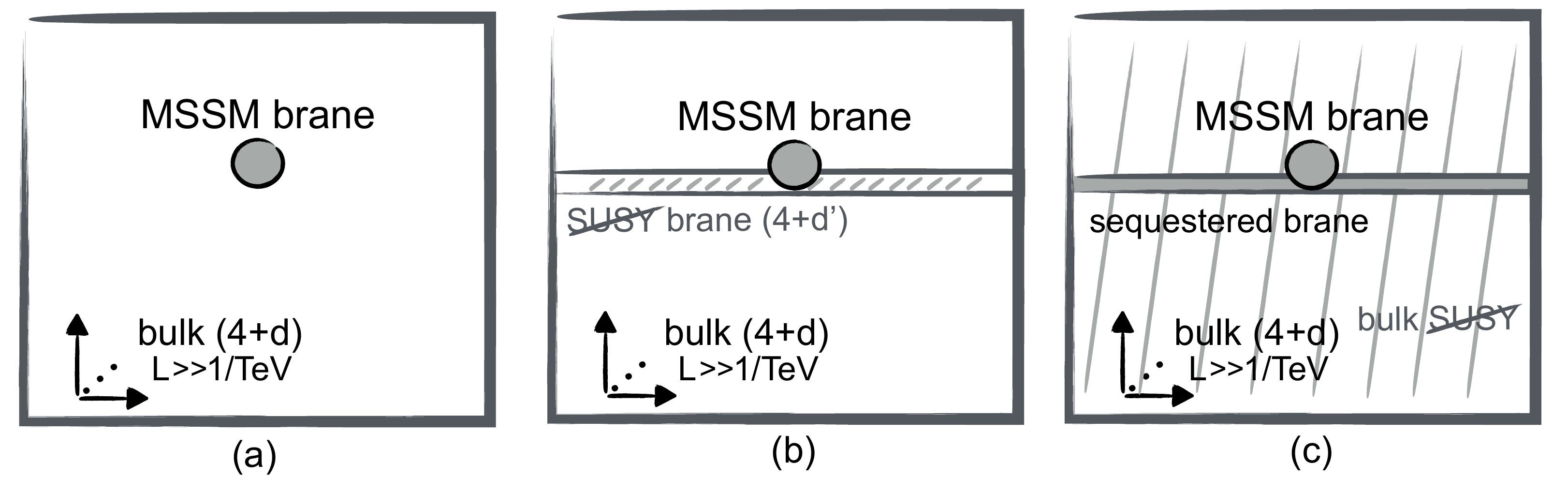}
  \caption{(a) The general set-up we consider, with the MSSM brane embedded in a large bulk with $d$  compact dimensions of size $L\gg\TeV^{-1}$. The MSSM brane may have structure at scales smaller than a $\TeV^{-1}$, and possible additional extra dimensions of size $\lesssim \TeV^{-1}$ are not depicted. (b) The same embedding, with the MSSM SUSY breaking shown explicitly to occur on a nearby brane extended in a $(4+d')$ dimensional subspace of the large bulk. (c) The same embedding of the MSSM, with SUSY breaking extended throughout the entire large bulk.  Additional states may live in the bulk or on sequestered branes of lower codimension as shown, and are candidates for light bulk LSPs. Although the SUSY breaking is present everywhere in the large bulk, it may be localized in further dimensions of size $\lesssim \TeV^{-1}$ not shown.}
    \label{fig:SUSYBreaking}
\end{figure}

The breaking of the supersymmetry remaining on the MSSM brane should be felt softly, giving superpartner masses $m_{\rm soft}\sim \TeV$. As depicted in Fig.~\ref{fig:SUSYBreaking}(b),
 the MSSM SUSY breaking can occur over any subspace of the bulk with dimension $4+d'$ ($d'\leq d$), leaving the SUSY breaking localized in the $d-d'$ transverse large dimensions. The goldstino degrees of freedom from this breaking will also propagate in $4+d'$ dimensions. After considering the mixing with the gravitino these degrees of freedom are lifted, but as we will show, for low SUSY breaking scales or $d' < d-2$ they can generically remain lighter than the MSSM states, providing a candidate for a bulk LSP. We discuss this case first, and then discuss other candidates for bulk LSPs which naturally occur in sequestered sectors and are particularly relevant when decays to the goldstino degrees of freedom become suppressed or kinematically inaccessible.

\subsection{Goldstino Bulk LSP}

Take the MSSM SUSY breaking to be parameterized by an F-term for a field localized on a brane in a $4+d'$ subspace of the bulk, $\langle F_{4+d'}\rangle$. In a theory without gravitational degrees of freedom, there is a massless goldstino propagating in $4+d'$ dimensions. First we study directly the decays of MSSM particles to this degree of freedom, and then we will discuss the effects of mixing the goldstino degree of freedom with the gravitino.

The couplings of the 3-brane localized MSSM states to the bulk goldstino $\eta$ can be inferred as usual from the soft masses, for example for couplings to a MSSM chiral multiplet $(f,\tilde{f})$,
\beq 
 \label{eq:Lgoldstino}
 \mathcal{L}_{\rm soft} = m^2_{f}f^\dagger f ~\delta(y) \rightarrow \sim \frac{m^2_{f}}{\langle  F_{4+d'} \rangle} f^\dagger \tilde{f} \eta ~ \delta(y) + h.c. 
\eeq
Note we use the canonical normalizations of a $(4+d')$-dimensional field for $\eta$ and $\langle  F_{4+d'} \rangle$.
In comparison to our earlier results for a bulk modulino Eq.~\ref{eq:SleptonModulino}, the decay $\tilde{e}_R \rightarrow e + \eta$ has the rate,
\beq
\label{eq:SleptonGoldstino}
\Gamma_{\rm tot} \approx \frac{M^{5+d'}}{8\pi \langle F_{4+d'}\rangle^2}\frac{\Omega_{d'}}{(2\pi)^{d'}}\int_0^1 x^{d'-1} (1-x^2)^2 dx.
\eeq
The distribution of KK masses in these decays is slightly softer than decays to a bulk modulino, and the overall rate depends on a higher power of $M$. This decay has the usual $1/\langle F \rangle^2$ rate expected for decays to a goldstino in 4d models.

An important consequence is that if SUSY breaking is localized on a 3-brane like the MSSM, then decays to the goldstino will not appear extra dimensional and will often dominate over decays to any bulk LSPs present. For example, even if SUSY breaking is at the fundamental scale, $\langle F_4 \rangle \sim M_*^2$, decays to the 4-dimensional goldstino are parametrically enhanced by powers of $\frac{M}{M_*}$  compared to decays to a modulino living in more than six dimensions. Thus decays to a bulk state with large codimension only occur generically when SUSY breaking is extended in the bulk. 

These results for the decays to a goldstino hold exactly in the limit that gravity decouples, $M_*\rightarrow \infty$ with $\langle F_{4+d'}\rangle$ held fixed. In the supergravity theory with finite $M_*$, the goldstino degree of freedom will mix with the gravitino; if there is a single source of SUSY breaking, the goldstino will be completely eaten by the gravitino, while if there is additional SUSY breaking elsewhere in the bulk some combinations will be left as a pseudo-goldstinos with perturbed mass spectra. 

The diagonalization of the full gravitino bulk+brane equations of motion and determination of the masses and couplings of gravitino KK modes is beyond the scope of this work, but fortunately the equivalent goldstino approximation provides good intuition, and further we expect it to provide accurate results for many scenarios. We can understand when the equivalent goldstino approximation remains valid by considering the locality of decays from the MSSM brane. A 3-dimensional MSSM state of mass $M$ will couple to gravitino states localized to a distance $\Delta \sim 1/M$ within the full bulk. We can therefore expect qualitatively correct results from considering only the light modes in the $4+d'$ dimensional theory after compactifying the $d-d'$  bulk dimensions tranverse to the SUSY breaking brane to a size $\sim 1/M$. If the $4+d'$ dimensions are approximately flat, there is a $4+d'$ dimensional gravitino mass term related to the mixing with the goldstino, 
\beq
m_{3/2,~(4+d')} \sim \frac{F_{4+d'}}{\sqrt{M_*^{2+d}/M^{d-d'}}}
\eeq
This mass sets the start of the gravitino KK tower in the 4d theory, and the equivalent goldstino approximation holds when $m_{3/2} \ll M$.

For a sufficiently small SUSY breaking scale $\langle F_{4+d'}\rangle$ this can always be satisfied, but a particularly interesting case is when the breaking is at the fundamental scale, $\langle F_{4+d'} \rangle \sim M^{2+d'/2}_*$. This occurs for example when the MSSM SUSY is broken by the presence of a nearby D-brane extended in $4+d'$ dimensions, with the $U(1)$ gaugino on the D-brane realizing the goldstino degree of freedom \cite{Bagger:1996wp,Dudas:2000nv,Antoniadis:2001pt,Pradisi:2001yv}. Then we find
\beq
 \frac{m_{3/2,~(4+d')}}{M} \sim \left(\frac{M}{M_*}\right)^{\frac{d-d'}{2}-1}.
\eeq
For $d' < d-2$, the perturbation is vanishing as $M_*\rightarrow\infty$ with brane soft masses $M$ fixed, and we expect the goldstino equivalence theorem to hold. We therefore expect our results for the decays to the goldstino degrees of freedom to be a good description of a wide class of models where the goldstino is the bulk LSP. On the other hand for the cases $d'=d-2$ and $d'=d-1$ the decays to all of the components of the gravitino become important unless the SUSY breaking scale is parametrically below $M_*$;  there may be interesting cases where such a gravitino is the LSP but these cases are subtle and left for future work (the super-higgs mechanism for the $d'=d-1$ case has been studied in 5d models in refs.~\cite{Bagger:2004rr,Benakli:2007zza}). In the case $d'=d$, the lightest gravitino KK mode can be heavier than the MSSM LOSP even for $\langle F_{4+d'}\rangle$  well below the fundamental scale. We discuss this case in the following section, focusing on decays to the variety of other motivated light bulk LSPs which may still arise even when the gravitino is heavy.


\subsection{Heavy Gravitino}
In models where SUSY breaking occurs throughout the entire large d-dimensional bulk, as depicted in Fig.~\ref{fig:SUSYBreaking}(c), the gravitino can obtain a large bulk mass lifting its lowest KK states to masses $\gtrsim m_{\rm soft}$ if the bulk SUSY breaking is communicated to the MSSM fields via gravitationally coupled operators and states, for instance via radion or dilaton mediation in additional $R \lesssim \TeV^{-1}$ sized extra dimensions. In such models, sequestered sectors with further suppressed soft masses propagating in $\leq d$ large dimensions can naturally occur, providing candidates for bulk LSPs.

\begin{figure}[t]
  \centering
  \includegraphics[width=.4\columnwidth]{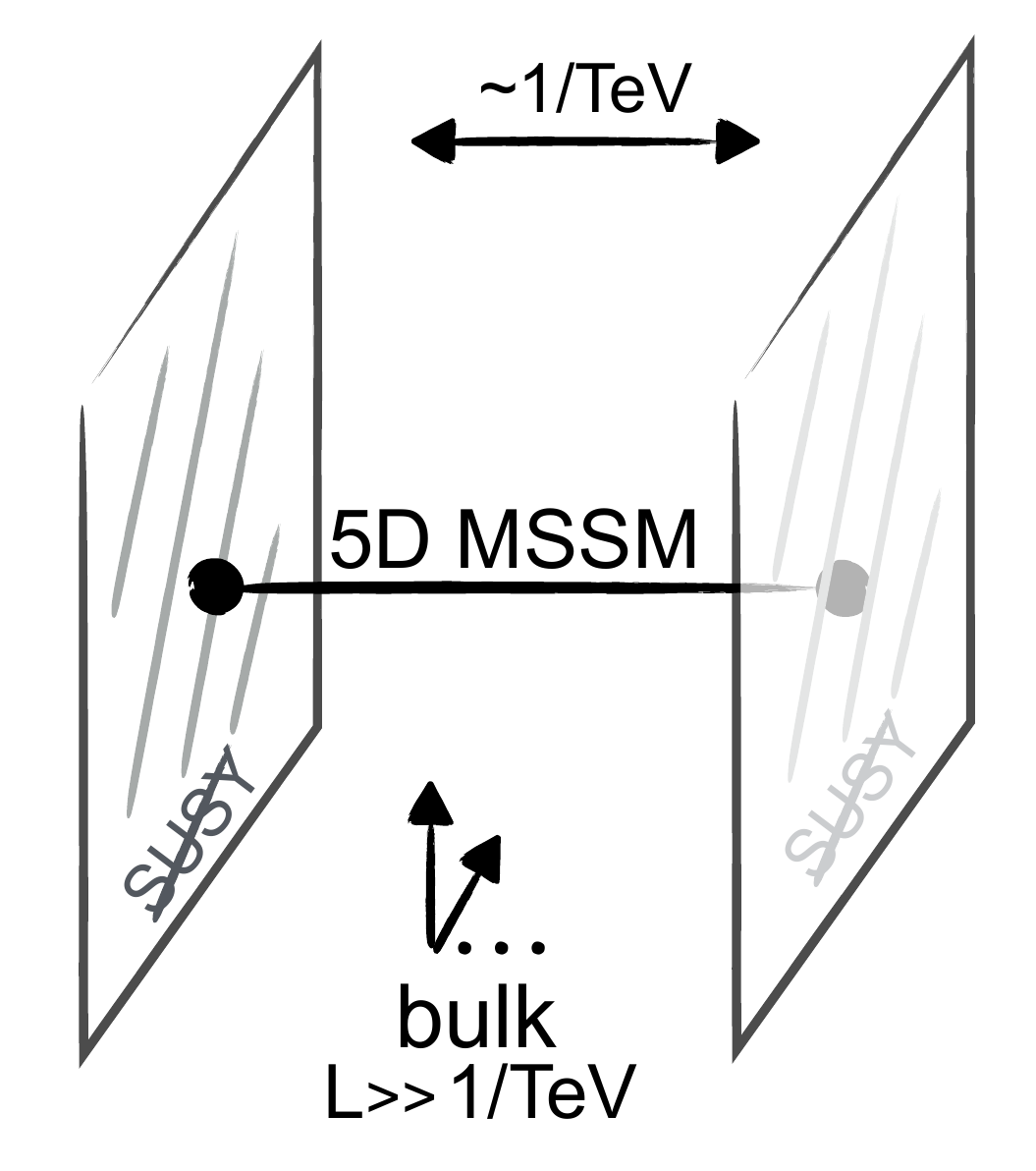}
  \caption{Embedding of 5d Scherk-Schwarz model in a $5+d$ dimensional theory as $\mathbb{R}^{3+1} \times (S_1 /\mathbb{Z_2} \times \mathbb{Z}'_2) \times \mathcal{M}_{d}$. The MSSM states live on 3-branes or 4-branes completely localized within the $d$ large compact bulk dimensions. The boundary conditions on each end of the $\TeV^{-1}$-sized dimension partially break the bulk supersymmetry, leading to a complete breaking of SUSY in the theory at scales below $\TeV$, with the breaking spread through the entire $4+d$ dimensional large bulk as in Fig.~\ref{fig:SUSYBreaking}(c) and giving large $\sim \TeV$ scale masses to the lightest gravitino KK modes. Additional states may live extended in the large bulk but localized at either endpoint of the $\TeV^{-1}$-sized dimension; they will be sequestered from the full SUSY breaking and can lead to a bulk LSP.}
    \label{fig:ScherkSchwarzBulk}
\end{figure}

As a simple example, we consider the class of models studies in refs. \cite{MNSUSY,Garcia:2014lfa,moreMNSUSY,Antoniadis:1998sd,Delgado:1998qr,Pomarol:1998sd,Barbieri:2003kn,Barbieri:2002sw,Barbieri:2002uk,Barbieri:2000vh,Marti:2002ar,Diego:2005mu,Diego:2006py}, where some of the MSSM states propagate in a $\sim \TeV^{-1}$ sized 5th dimension given by a segment $S_1 /\mathbb{Z_2} \times \mathbb{Z}'_2$ which breaks SUSY by Scherk-Schwarz boundary conditions \cite{Scherk:1978ta,Scherk:1979zr}. If this is embedded trivially in a large gravitational bulk as  $\mathbb{R}^{3+1} \times (S_1 /\mathbb{Z_2} \times \mathbb{Z}'_2) \times \mathcal{M}_{d}$, as illustrated in Fig.~\ref{fig:ScherkSchwarzBulk}, then the SUSY breaking boundary conditions on $S_1$ give a d-dimensional mass $\sim \TeV$ to the gravitino uniformly in the large bulk and generates soft masses $\sim \TeV$ for the MSSM states. This can equivalently be described as radion mediation with $F_{T,4+d} \sim \TeV \times M_*^{1+d/2}$ \cite{Ferrara:1988jx,Porrati:1989jk,Marti:2001iw,Kaplan:2001cg,Gherghetta:2001sa}. Additional states living on branes localized in the 5th dimension and extended in the bulk dimensions obtain soft masses through the gravitational couplings only at loop level \cite{Gherghetta:2001sa}, and can naturally arise as the bulk LSP. This scenario is illustrated in Figure~\ref{fig:ScherkSchwarzBulk}.

The bulk LSP in this scenario can have a variety of forms. We have all ready discussed in detail the coupling to a bulk modulus field, which may arise for example due to branes wrapping additional $M_*^{-1}$ sized dimensions or cycles. Other motivated possibilities for bulk LSPs living on a sequestered brane include a $U(1)'$ gaugino, a bulk axino, or a bulk sneutrino.


\subsubsection{Bulk Axino}
If the strong CP problem is solved by the axion in a model with a low fundamental scale, then the axion multiplet must propagate in some of the bulk dimensions. A simple possibility for the form of the effective couplings of the axino to the chiral multiplets of the MSSM is
\beq
\label{eq:Laxino}
\mathcal{L}_{\rm axino} = \delta(y)\frac{c_{f1}}{f_*^{(d'+2)/2}} \tilde{a}f \tilde{f}^{\dagger} + \delta(y)\frac{c_{f2}}{f_*^{(d'+2)/2}} \tilde{a}f \tilde{f^c}  + h.c.
\eeq
where $c_{f1}\sim(10^{-3}-10^{-4}) M_{\tilde{g}}$ is radiatively generated from the anomaly couplings \cite{Covi:2002vw} and $c_{f2}\sim m_f$ is generated if the Higges are charged under the PQ symmetry (DFSZ axion) \cite{Bae:2011iw,Baryakhtar:2013wy}.  More general forms of the axion supermultiplet couplings to the visible-sector fields \cite{Bae:2011jb,Baryakhtar:2013wy} lead to qualitatively similar results. The two-body decays mediated by the interactions of Eq.(\ref{eq:Laxino})  dominate over 3-body decays through off-shell gauginos \cite{Covi:2002vw} and  can easily dominate over rates into  gravitationally coupled states if scales are chosen to give a (3+1)-dimensional axion scale $f_a\lesssim10^{16}~\GeV$.

\subsubsection{Bulk $U(1)'$ gaugino}

Another interesting candidate for a bulk LSP is the gaugino of a bulk $U(1)'$ coupling to the MSSM fields through the $B-L$ current or kinetic mixings \cite{Abel:2003ue, Abel:2008ai}. In terms of a dimensionless gauge coupling, $\tilde{g}$, the couplings to the chiral brane fields are of the form,
\beq
\label{eq:Lgaugino}
\mathcal{L}_{\rm gaugino} = \delta(y)\frac{\tilde{g}}{M_*^{d'/2}} \tilde{\lambda}f \tilde{f}^{\dagger} + h.c.
\eeq
Note that this coupling is of lower dimension than the decays to gravitational states or an axino, and naturally can dominate over other channels if present. The $U(1)'$ can be broken supersymmetrically on the 3-brane or elsewhere in the bulk to evade constraints on the gauge bosons. Limits on the scale $M_*$ from direct single production of bulk $KK$ gauge bosons will be enhanced compared to the KK graviton limits, while limits from contact operators will not be substantially changed.

\subsubsection{Bulk sneutrino}

A final well-motivated bulk LSP candidate is a sneutrino superpartner of one of the sterile neutrinos that can exist in the extra-dimensional bulk.  Such bulk sterile neutrinos
can explain the observed neutrino masses and mixings by way of a volume-enhanced effective (3+1)-dimensional Majorana mass leading to a
see-saw like formula \cite{ArkaniHamed:1998vp}.   The details of the brane-bulk couplings, and particularly their flavour structure, are potentially interesting in the
regime discussed in Section 4 probing extra dimensions, as they may give additional clues to the structure of the underlying neutrino model.

\subsection{Summary}
\label{sec:BulkLSPSummary}

We have discussed a variety of motivated possibilities for decays to a bulk LSP: a goldstino of SUSY breaking extended in the bulk, a modulino, a $U(1)'$ gaugino, and an axino. We have focused on the case of the decays of a massless sfermion, but we can more generally consider the decay of any MSSM LOSP to a bulk LSP.

\begin{figure}[t]
  \centering
  \includegraphics[width=.7\columnwidth]{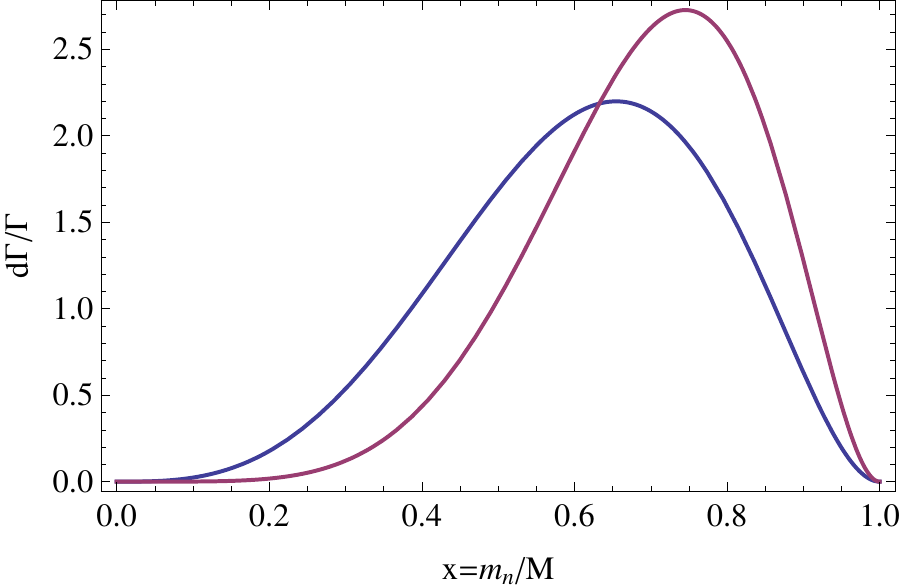}
  \caption{Differential distribution of KK masses for the decay $\tilde{e}_R \rightarrow e + {\tilde X}$ for differing candidate bulk LSPs, ${\tilde X}$, ${\tilde X}=\eta$, a bulk goldstino (leftmost, blue curve), or, ${\tilde X}=\psi$, a bulk
  modulino (rightmost, red curve). A bulk $U(1)'$ gaugino or axino has the same distribution as the goldstino. For both cases we fix $d=6$. This illustrates that auto-concealment is slightly more effective for the modulino coupling than other bulk LSPs.}
    \label{fig:SleptonGravitino}
\end{figure}

\begin{figure}[t]
  \centering
  \includegraphics[width=.7\columnwidth]{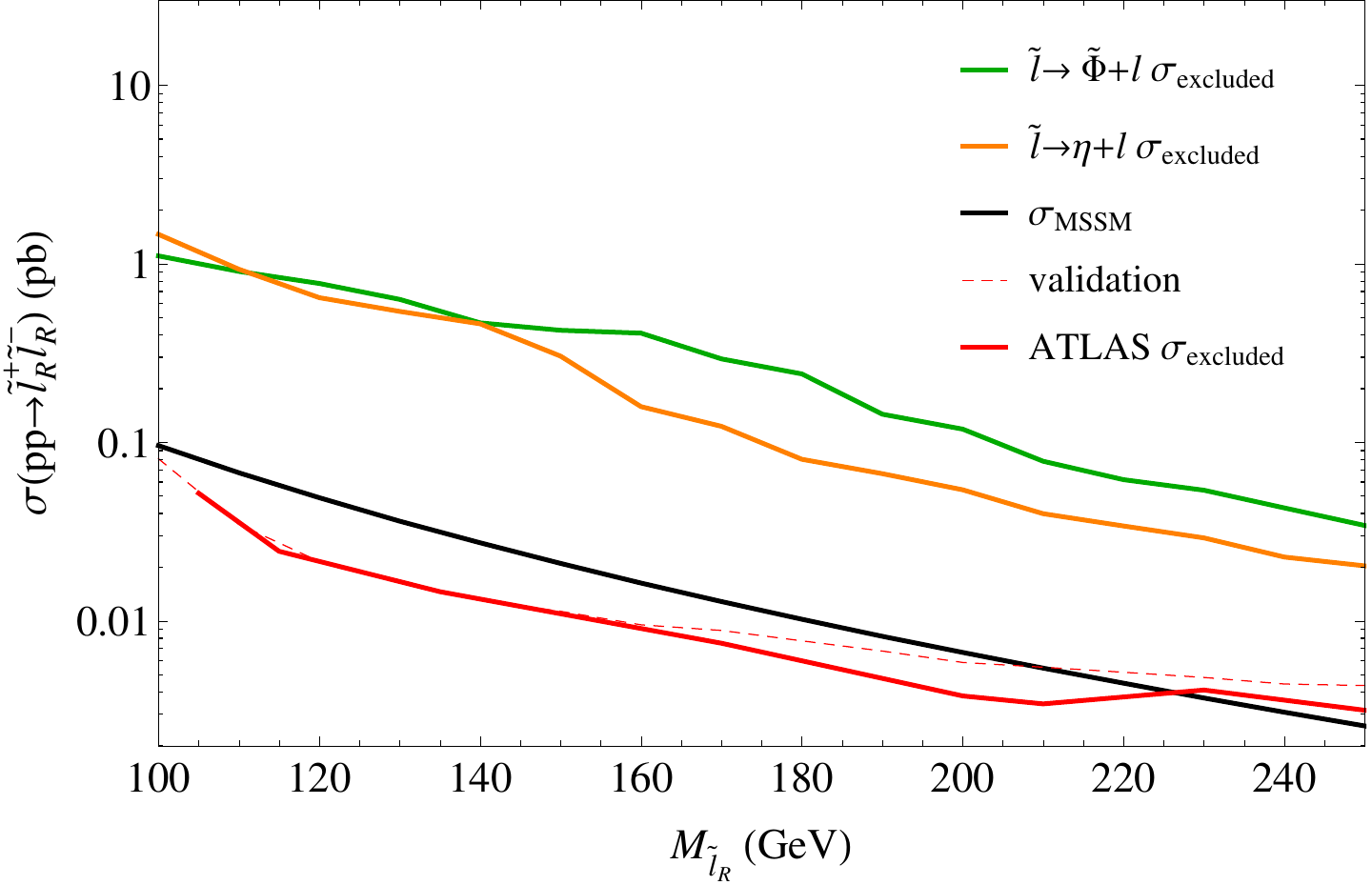}
  \caption{Strongest upper bound on degenerate ${\tilde \mu_R}$, ${\tilde e_R}$ pair production cross sections from ATLAS $l^+ l^- + \MET$  $m_{T2}$ \cite{ATLAS-CONF-2013-049} and razor analyses \cite{ATLAS-CONF-2013-089}.  A monojet search was also considered \cite{ATLAS-CONF-2012-147} but sets weaker limits.  The top two curves corresponds to sleptons promptly decaying to the KK tower of a modulino ($\psi$, green) or goldstino ($\eta$, orange),  in $d=6$ extra dimensions.  The $m_{T2}$ analysis is the more effective of the two searches except below 170 and 150 GeV for decays to bulk modulinos and goldstino respectively.  Solid red (lowest) curve gives the observed ATLAS upper bound on the RH slepton production cross section from \cite{ATLAS-CONF-2013-049} for decays to a massless LSP. For validation, the dashed red curve gives the same bound using our simulation. Black curve gives the predicted NLO direct production cross section \cite{Beenakker:1999xh} illustrating that RH sleptons are excluded up to $\sim 225~\GeV$ for a single massless LSP.  For the searches considered both the modulino and goldstino eliminate present limits on direct RH slepton production.}
\label{fig:SleptonAllLSPLimits}
\end{figure}

\begin{table}[h]\centering
\begin{tabular}{c|c|c|c}
                                   					& $\psi$ 					& $\eta/\tilde{a}$ 			& $\tilde{\lambda}'$  	\\ 
									& $(\alpha,\beta,\delta)$		& $(\alpha,\beta,\delta)$		& $(\alpha,\beta,\delta)$		\\ \hline
$\tilde{f}_{L/R} \rightarrow f + \tilde{X}$   	& $(2,2,0)$       			& $(0,2,0)$				& $(0,2,0)	$		 	\\ \hline
$\tilde{H}\rightarrow h/Z + \tilde{X}$ 		& $(0,2,1)$  				& $(0,2,1)$           			& $(0,2,1)	$			\\ \hline
$\tilde{\lambda}\rightarrow V/g + \tilde{X}$ & $(0,3,0)$				& $(0,3,0)$           			& --  				\\ \hline
\end{tabular}
\caption{Possibilities for the kinematic distribution, Eq.(\ref{eq:dGammadX}), in the decays of sfermion ($\tilde{f}$), higgsino-like ($\tilde{H}$), and wino/bino/gluino-like LOSPs ($\tilde{\lambda}$) to variety of bulk LSPs $\tilde{X}$: a modulino $\psi$, Eqs.(\ref{eq:Lmodulino}-\ref{eq:SleptonModulino}), a goldstino $\eta$, Eqs.(\ref{eq:Lgoldstino}-\ref{eq:SleptonGoldstino}), an axino $\tilde{a}$,
Eq.(\ref{eq:Laxino}), a $U(1)'$ gaugino $\tilde{\lambda}'$, Eq.(\ref{eq:Lgaugino}). These distributions hold in the massless limit for the SM particle in the decay, $(M-m_n)\gg m_h, m_V, m_t$.}
\label{tab:alphabeta}
\end{table}

From an observational point of view, these various possibilities for bulk LSPs can be simply summarized. The general form for the differential distribution of KK masses in the decay of a brane-localized LOSP to a bulk LSP and a massless SM particle is a sum of terms of the form
\beq
\label{eq:dGammadX}
\frac{d\Gamma}{dx} \sim x^{d-1+\alpha} (1-x^2)^\beta(1+x^2)^\delta;\;\;\;x\equiv m_n/M,
\eeq
where the $(1-x^2)^\beta$ factor captures the 3D phase space dependence, and the $x^{d-1+\alpha}$ factor includes the bulk phase space factor. The remaining freedom in $\alpha,\beta,\delta$ comes from the matrix element for a given process. Typically a single term of the form Eq.(\ref{eq:dGammadX}) dominates the distribution; qualitatively, $\alpha$ and $\beta$ are the most important for determining the shape of the distribution because of their zeros -- a large $\alpha$ and a small $\beta$ corresponds to a distribution of decays peaked at the largest kinematically allowed KK masses.  Table~\ref{tab:alphabeta} surveys the forms of the distribution for a variety of combinations of brane LOSP and bulk LSP candidates decaying to an effectively massless SM state. The decays to a goldstino, axino, and $U(1)'$ gaugino have similar matrix elements and all follow the same kinematic form. They do not share the helicity suppression of decays to lighter KK modes found for the modulino-like coupling Eq.~\ref{eq:Lmodulino}, leading to distributions slightly softer than decays to a modulino. 

Figure~\ref{fig:SleptonGravitino} compares the distributions of KK masses for the decays of a RH slepton (${\tilde \mu_R},{\tilde e_R}$) to a goldstino or modulino LSP, and Figure~\ref{fig:SleptonAllLSPLimits} compares the corresponding effects on experimental limits in slepton searches.

\section{Conclusions}

We have presented a mechanism---auto-concealment in extra dimensions---which significantly weakens present search limits for some SUSY models. Auto-concealment applies to theories wherein the LOSP is a brane localized state while the LSP is a bulk state, producing a dense KK tower of LSP excitations with increasing mass, $m_n$, that automatically brackets the LOSP mass without further tuning.  The increased density of states at higher mass due to the bulk phase space factor $\sim m_n^{d-1}$ favours LOSP decays to the heaviest KK states, dynamically generating a quasi-compressed spectra, as discussed in Section~\ref{sec:couplingsdecays} and shown in Figures~\ref{fig:basicidea} and~\ref{fig:SleptonModulino}. If the scale $M_*$ is such that decays from the LOSP to the LSP are prompt, typical handles used in SUSY searches such as visible energy and $\MET$ are then dynamically suppressed as we discussed in Section~\ref{sec:collider}.  This reduces both $\MET$ and visible energy in SUSY events (unlike R-parity violation for example, which increases visible energy). 

Auto-concealment can occur for a variety of visible-sector LOSP candidates.  
In particular, we find that LHC limits on right-handed slepton LOSPs evaporate in the case of prompt decays to a bulk modulino (see Figure~\ref{fig:sleptonxsplot}), while the LHC limits on stop LOSPs weakens to $\sim350\div410~\GeV$ (see Figure~\ref{fig:stopxsplot}). 
Present LHC limits on direct production of degenerate first and second generation squarks similarly drop to $\sim 450~\GeV$ (see Figure~\ref{fig:squarkxsplot}).

As discussed in Section~\ref{sec:collider} the mechanism is effective for a variety of visible-sector superpartner spectra, but not for all kinds.  In particular, auto-concealment does not
significantly weaken limits driven by deep cascade decays to the LOSP.  In addition, the mechanism is most effective when the bulk LSP propagates in $>2$ large extra dimensions. As discussed in detail in Section~\ref{sec:varieties}, decays to motivated bulk states like axinos, $U(1)'$ gauginos, and modulinos propagating in many extra dimensions can be dominant when SUSY breaking also extends in some of the large bulk directions, and the goldstino itself can also be an attractive bulk LSP candidate.
Thus we find that a wide variety of visible-sector SUSY spectra, LOSP candidates, and relevant bulk sparticle states lead to efficient auto-concealment.
The auto-concealment mechanism also applies to more than just SUSY theories, broadly speaking to any theory wherein a discrete quantum number is shared
between brane and bulk states and where the analog of the LSP is a bulk state.

Though discovery becomes more difficult for SUSY spectra without common deep cascade decays to the LOSP, it is not impossible.  For the examples of stops and squarks of the first two generations we found searches designed to pick out compressed spectra \cite{Aad:2014qaa} and monojets \cite{ATLAS-CONF-2012-147} remain effective, albeit at lower masses than traditional searches.  For the case of sleptons, searches modified to keep lower-energy leptons in the signal region \cite{Buckley:2013kua,Rolbiecki:2012gn,Giudice:2010wb,Schwaller:2013baa,Baer:2014kya} could be useful.  

If superpartners are eventually discovered then observations of LOSP decays may prove to be a powerful window into extra dimensions.  This is especially true if the scale $M_*$ is large enough to lead to LOSP tracks in the detector, displaced vertices, or stopped out-of-time decays.  As we discussed in Section~\ref{sec:probing}
searches for such signals could probe the underlying gravitational mass scale up to $M_*\lsim 10^9~\GeV$!

Finally, we caution the reader that although we expect the limits obtained from validated CheckMATE analyses to provide a good estimate of the best current limits from the LHC experiments, there may be increased uncertainty because these searches are particularly sensitive to the tails of kinematic distributions and some of the most recent analysis updates are not presently available. We hope that this work motivates further study of these signals by the experimental collaborations.


\subsection*{Acknowledgments}
We thank Asimina Arvanitaki, Masha Baryakhtar, Mireia Crispin-Ortuzar, Alexandru Dafinca, Claire Gwenlan, Junwu Huang, William Kalderon, William Fawcett, Juan Rojo, and Giovanni Villadoro for useful discussions, and especially Alan Barr for very helpful discussions on experimental considerations.  We thank Jonathan Patterson for setting us up on the Oxford Theory Computing Cluster.   We also wish to thank the CERN Theory Group for hospitality during portions of this work. This work was partially supported by ERC grant BSMOXFORD no. 228169 and by NSF grant PHY-1316706. JMR is grateful to support from the STFC (UK). KH is supported by an NSF Graduate Research Fellowship under Grant number DGE-0645962 and by the US DoE under contract DE-AC02-76SF00515. JS gratefully acknowledges support from the United States Air Force Institute of Technology. The views expressed in this letter are those of the author and do not reflect the official policy or position of the United States Air Force, Department of Defense, or the US Government.

\footnotesize
\bibliography{AutoConcealment}
\end{document}